%% file: MainOnly.tex
\documentclass[fleqn,10pt]{wlscirep}
\usepackage[utf8]{inputenc}
\usepackage{subfigure,xr}
\usepackage{amsmath,amsfonts}	
\newcommand{\Ei}{{\bf E}_{i}}
\newcommand{\eig}{\varepsilon_\alpha}
\newcommand{\eigw}{{\bf w}_\alpha}

\newcommand{\SqrtTE}{\sqrt{\varepsilon^{\texttt{TE}|j}_{nl}}}
\newcommand{\SqrtTM}{\sqrt{\varepsilon^{\texttt{TM}|j}_{nl}}}

\newcommand{\eigTMl}{\varepsilon^{\texttt{TM}|j}}
\newcommand{\eigTEl}{\varepsilon^{\texttt{TE}|j}}

\newcommand{\re}[1]{\mbox{Re}\left\{#1\right\}}

\newcommand{\im}[1]{\mbox{Im}\left\{#1\right\}}

\newcommand*\CC[1]{{\bf d}_{#1}}

\newcommand*\CCo[1]{{\bf d}_{#1}^\parallel}
\newcommand*\CCp[1]{{\bf d}_{#1}^\perp}

\newcommand*\CoeffdTE[1]{h_{#1}^{\nu|\texttt{TE}|j}}
\newcommand*\CoeffdTM[1]{h_{#1}^{\nu|\texttt{TM}|j}}

\newcommand*\CoeffdTEmet[1]{h_{#1}^{\texttt{TE}|j}}
\newcommand*\CoeffdTMmet[1]{h_{#1}^{\texttt{TM}|j}}

\newcommand*\CoeffhTE[1]{\tilde{H}_{#1}^{\nu|\texttt{TE}|j}}
\newcommand*\CoeffhTM[1]{\tilde{H}_{#1}^{\nu|\texttt{TM}|j}}

\newcommand*\CoeffhhTE[1]{H_{#1}^{\nu|\texttt{TE}|j}}
\newcommand*\CoeffhhTM[1]{H_{#1}^{\nu|\texttt{TM}|j}}

\newcommand*\CoeffhhhTM[2]{H_{#1}^{\nu|\texttt{TM}|#2}}
\newcommand*\CoeffhhhTE[2]{H_{#1}^{\nu|\texttt{TE}|#2}}

\newcommand*\CoeffhhhhTM[1]{H_{#1}^{\nu|\texttt{TM}}}
\newcommand*\CoeffhhhhTE[1]{H_{#1}^{\nu|\texttt{TE}}}

\newcommand*\CoefflTM[1]{H_{#1}^{\parallel \, |\texttt{TM}}}

\newcommand*\CoefftTM[1]{H_{#1}^{\perp \, |\texttt{TM}}}
\newcommand*\CoefftTE[1]{H_{#1}^{\perp \, |\texttt{TE}}}

\newcommand*\EigDimer[1]{ {\varepsilon}_{#1}}

\newcommand*\Cah[1]{{\bf e}_{#1}^{\texttt{TM}|j}}
\newcommand*\Cbh[1]{{\bf e}_{#1}^{\texttt{TE}|j}}

\newcommand*\Ca[1]{{\bf e}_{#1}^{\texttt{TM}}}
\newcommand*\Cb[1]{{\bf e}_{#1}^{\texttt{TE}}}
\newcommand{\Mi}[3]{{\bf M}_{#3} ^{\left(#1\right)} \left( #2 \right)}
\newcommand{\Ni}[3]{{\bf N}_{#3}^{\left(#1\right)} \left( #2 \right)}
\newcommand*\C[1]{{\bf C}^{\left(#1\right)}}
\newcommand{\rbar}{{\bf r}_{\bar{j}}}
\newcommand{\rj}{{\bf r}_{j}}

\newcommand*\Caj[1]{{\bf e}_{#1}^{\texttt{TM}|j}}
\newcommand*\Cbj[1]{{\bf e}_{#1}^{\texttt{TE}|j}}

\newcommand{\rvers}{\hat{\bf r}}

\newcommand{\J}[2]{j_{#2}\left(#1\right)}

\newcommand{\Es}{{\bf E}_{S}}

\newcommand{\inoutTM}[2]{\sigma_{#1}^{\mathtt{TM}|#2}}
\newcommand{\inoutTE}[2]{\sigma_{#1}^{\mathtt{TE}|#2}}

\newcommand{\Nmax}{N_{max}}
\newcommand{\Lmax}{L_{max}}
\newcommand{\uTM}[2]{u_{#1}^{\mathtt{TM}|#2}}
\newcommand{\uTE}[2]{u_{#1}^{\mathtt{TE}|#2}}

\newcommand{\vecD}{{\bf y}_{pmq}}

\newcommand*\Proj[2]{\mathcal{P}^{\mathtt{#1}|j}_{pmnl} \left\{#2\right\}}

\newcommand{\NN}[2]{{\bf N}_{#1}^{\left( #2 \right)}}
\newcommand{\MM}[2]{{\bf M}_{#1}^{\left( #2 \right)}}

\title{Full-wave electromagnetic modes and hybridization  in nanoparticle dimers}

\author[1]{Mariano Pascale}
\author[1]{Giovanni Miano}
\author[1,2]{Roberto Tricarico}
\author[1,*]{Carlo Forestiere}
\affil[1]{Department of Electrical Engineering and Information Technology, Universit\`{a} degli Studi di Napoli Federico II, via Claudio 21, Napoli, 80125, Italy}
\affil[2]{ICFO Institut de Ciències Fotòniques, The Barcelona Institute of Science and Technology, 08860 Castelldefels, Barcelona, Spain}

\affil[*]{carlo.forestiere@unina.it}


\begin{abstract}
The plasmon hybridization theory is based on a quasi-electrostatic approximation of the Maxwell's equations. It does not take into account magnetic interactions, retardation effects, and radiation losses. Magnetic interactions play a dominant role in the scattering from dielectric nanoparticles. The retardation effects play a fundamental role in the coupling of the modes with the incident radiation and  in determining their radiative strength; their exclusion may lead to erroneous predictions of the excited modes and of the scattered power spectra. Radiation losses may lead to a significant broadening of the  scattering resonances. We propose a hybridization theory for non-hermitian composite systems based on the full-Maxwell equations that, overcoming all the limitations of the plasmon hybridization theory, unlocks the description of dielectric dimers. As an example, we decompose the scattered field from silicon and silver dimers, under different excitation conditions and gap-sizes, in terms of dimer modes, pinpointing the hybridizing isolated-sphere modes behind them.
\end{abstract}
\begin{document}

\flushbottom
\maketitle
%
%
\thispagestyle{empty}

\section*{Introduction}
\input{Source/Intro}
\section*{Modal Decomposition of Electromagnetic Scattering}
\input{Formulation/Formulation}
\section*{Results and Discussion}
We investigate the hybridization mechanism in silver and silicon dimers for the modes that are excited in the scattering under longitudinal and transverse plane-wave excitations.
Specifically, we consider two spheres with the same radius, namely an homo-dimer, with $R_1 = R_2 = R$ ($x  =x_1= x_2$) and edge-edge separations of $R/4$, $R$, and $\infty$, corresponding to a center-center separation of $D = 9/4 R$, $D = 3 R$, $D = \infty$. We vary the size parameter $x$ in the interval $\left[0.6, 1.7\right]$. For different separations the set of dimer-modes varies, nevertheless the set of considered isolated-sphere mode used for their representation is the same. This is the advantage of the hybridization approach. The total number of considered dimer-modes is $N=1600$, with $q=1,\ldots,320$, $m = 0,\ldots,4$, $p=e$.  Nevertheless,  in each of the four considered scenarios, the scattering is dominated by different subsets of dimer-modes whose number is much less than $N$. 
The isolated-sphere modes used to represent the dimer modes are $ \left\{ {\bf e}_{emnl}^\mathtt{TM}, {\bf e}_{omnl}^\mathtt{TE} \right\}$ with $n=1 \ldots 8$, $l=1 \ldots 10$, and $m=1\ldots 4$. 
\input{AgZ_R4/AgZ}
\input{AgX_R4/AgX}
\input{SiZ_R4/SiZ}
\input{SiX_R4/SiX}
	
\section*{Conclusions}
\input{Source/Conclusions}

\section*{Methods}
\input{Source/Methods}
\setcounter{figure}{0}
\setcounter{table}{0}
\setcounter{equation}{0}

\renewcommand\theequation{S\arabic{equation}}    
\renewcommand\thefigure{S\arabic{figure}}    
\renewcommand\thetable{S\arabic{table}}    


\input{MainOnly.bbl}
\section*{Author contributions statement}
C.F. and G.M. conceived the idea,  M.P. and C.F developed the theoretical formulation. M.P wrote the numerical codes and conducted the numerical experiments. All authors analysed the results and reviewed the manuscript. 

\section*{Additional Information}
{\bf Competing interests:} The authors declare no competing interests.

\end{document}

%% file: Source/Intro.tex
The description of the electromagnetic scattering from nanostructures in terms of their resonances and modes is essential for both the analysis and the engineering of the field-matter interaction. Compared to the direct solution of the scattering problem, the description in terms of resonances and modes, which solely depend on the inherent properties of the nanostructure, i) offers intuitive insights into the physics of the problem; ii) enables the rigorous understanding of interference phenomena, including Fano resonances, in terms of the interplay among well-identified modes; iii) suggests how to shape the excitation to achieve an assigned electromagnetic response.  

In closed electromagnetic systems the
definition of resonances and modes is straightforward \cite{hanson2013operator}. On the contrary, in open systems, where the electromagnetic field occupies an unbounded domain, this definition is challenging. In an oversimplified but widespread approach, the electromagnetic resonances of a body are found as the peaks of its scattered power spectrum when a frequency-tunable probe field illuminates it. The corresponding electric field distributions are denoted as ``modes''. This approach
is flawed because: it hides the modes that cannot be excited by the chosen incident field; it disregards the fact that a peak can be due to the interplay of two or more modes; it does not help to interpret interference phenomena. 
Several more rigorous approaches are possible, grounded in different choices of the modes. The quasi-normal modes \cite{ching1998quasinormal,kristensen2013,lalanne2018light}
and the characteristic modes \cite{Garbacz} are widely used to study open systems. The quasi-normal modes depend on both the material and the geometry of the scatterer. They are not orthogonal in the usual sense and they diverge exponentially at large distances \cite{kristensen2013}, thus they need to be normalized \cite{sauvan2013theory,kristensen2015normalization}. The characteristic modes are increasingly used in nanophotonics \cite{makitalo2014modes}. They are real and satisfy a weighted orthogonality. They depend on the frequency, on the geometry and on the material composition of the scatterer. A third choice, that we embrace throughout this work, is represented by the material-independent modes \cite{Bergman80,Forestiere16}. The material-independent modes allow separating the role of geometry, material, and incident electromagnetic field; thus provide fundamental information on the resonant electromagnetic behaviour of bodies
that other approaches hide. They are not orthogonal in the usual sense but, unlike quasi-normal modes, they satisfy the radiation condition at infinity.

Unfortunately, regardless of the chosen definition, the electromagnetic modes of composite electromagnetic structures can be extremely complicated. They exhibit a complex dependence not only on the geometry of the constituent parts but also on their spatial arrangement. As a consequence, any significant change of the mutual arrangement of the constituent parts usually results in a meaningful change of the resulting modes. This fact undermines the intuitive understanding of the physics of the overall system.

Interacting nanoparticles, which can be either metallic or dielectric, constitute an example of composite open electromagnetic system. In particular, interacting metal nanoparticles have been extensively studied \cite{maier2010plasmonics}: they exhibit greater electric field enhancement with respect to their isolated counterparts \cite{schuller2010plasmonics}, they show tunability of the resonance position \cite{halas2011plasmons,prodan2003structural} and of the scattering directionality as the interparticle spacing varies, and novel physical properties, such as Fano resonances \cite{luk2010fano,kamenetskii2018fano,forestiere2013theory}. 
The plasmon hybridization theory \cite{prodan2003hybridization,nordlander2004plasmon} has been a cornerstone for the modelling of such metal systems. The plasmon hybridization consists in the representation of the modes of a complex plasmonic nanostructure, i.e. the ``plasmonic molecule'', in terms of the modes of its constituent parts, i.e. the ``plasmonic atoms''. Thus, even if the mutual spatial arrangement of the atoms is changed, the modes of the plasmonic molecule are represented in terms of the {\it same} set of atomic modes, while only the atomic modes weights change. Many studies have demonstrated that very complex molecular modes arise from the hybridization of just few atomic modes  \cite{Wang06,nanostar07,Oligomer,brandl2005plasmon,nordlander2004plasmon}.
%
However, the theory of plasmon hybridization \cite{prodan2003hybridization} is grounded in the electrostatic theory: it is based on compact hermitian operators and orthogonal electric field modes. This theory is only applicable to metal structures much smaller than the incident wavelength, because  the magnetic interactions and the radiation effects are absent. In other words, the plasmon resonators are treated as they effectively were {\it closed} resonators. For this reason, although its validity domain can be extended to include weak radiative contributions by using perturbation approaches \cite{Mayergoyz05} or by adding retardation to the  Coulomb potential \cite{turner2010effects}, it completely fails to describe dielectric resonators, which are dominated by magnetic interactions \cite{kuznetsov2012magnetic}.

Recently, dielectric resonators are gaining increasing attention in nanotechnology and many researchers currently suggest that high index dielectrics may be a cheaper alternative to noble metals for a variety of applications \cite{Evlyukhin:10,Evlyukhin:11,garcia2011strong,Kuznetsovaag2472,doi:10.1021/acsphotonics.7b01038}. 
This interest is motivated by the evidence that the enhancement of electric and magnetic fields in high-index nanostructures is of the same order of magnitude of the one achievable in metal nanostructures. Furthermore, the physics governing the scattering from high index dielectric nanoparticles is far richer than the physics behind the scattering from metal nanoparticles, due to the possibility of exciting magnetic modes  \cite{evlyukhin2012demonstration,kuznetsov2012magnetic,vandeGroep:16,kapitanova2017giant} and due to the presence of multimode interference, which may lead to the formation of Fano-resonances \cite{luk2010fano,Forestiere2018}. The resonances and modes and the corresponding properties of an isolated-sphere  \cite{Forestiere16,forestiere2017nanoparticle} and of a coated sphere \cite{pascale2017spectral} have been recently studied in the full-Maxwell regime.

Dimers of dielectric particles coupled in the near field zone may exhibit significant enhancement of both electric and magnetic fields \cite{Sigalas:07,albella13,bakker2015magnetic,
zywietz2015electromagnetic,boudarham2014enhancing,
mirzaei2015electric} with 
reduced heat conversion, \cite{albella2014electric,caldarola2015non}
directional Fano-resonances \cite{yan2015directional},
and strong directional scattering \cite{albella2015switchable,wang2014broadband}. A theory of hybridization in Si dimers, although limited to electric and magnetic dipole-dipole interactions, have been theoretically proposed in \cite{albella13} and experimentally validated in \cite{zywietz2015electromagnetic}. More recently, the hybridization in Si \cite{vandeGroep:16} and AlGaAs \cite{mcpolin2018imaging} dimers has been experimentally studied.
 
In this paper, we derive the resonances and modes of a sphere dimer by using the  full-Maxwell equations. We describe the dimer-modes in terms of the hybridization of the modes of the two constituent spheres: each dimer-mode is expressed in terms of a weighted linear combination of a set of isolated-sphere modes. The mathematical problem thus becomes a system of linear algebraic equations for the expansion coefficients. This scheme has been proposed by Bergman and Stroud \cite{Bergman80} in 1980. However, they  applied this method only in the long-wavelength limit when all the radii, as well as the interparticle separations, are small compared to the wavelength outside the scatterers. The approach we propose applies to both plasmonic and dielectric dimers regardless of their size. This fact enables us to address, for the first time, the mode analysis and the hybridization in silicon dimers in the full-Maxwell regime, and to refine the understanding of plasmon-mode hybridization in a full-wave scenario. 


%% file: Formulation/Formulation.tex
Here, we summarize the description in terms of the material-independent modes \cite{Forestiere16} of the full-wave electromagnetic scattering by a body which occupies a domain $\Omega$ of characteristic dimension $l_c$ suspended in vacuum and excited by a time harmonic electromagnetic field with frequency $\omega$ incoming from infinity, i.e. $\re{ {\bf E}_i } = \re{ {\bf E}_0 e^{- i \omega t} }$. We define the normalized size parameter of the body as $x=2\pi l_c / \lambda$, where $\lambda$ is the vacuum wavelength $\lambda = 2 \pi c / \omega$. The body is made by a linear material, it is also assumed to be non-magnetic, isotropic, homogeneous in time and space, non dispersive in space, and time-dispersive with relative permittivity $\varepsilon_R \left( \omega \right)$. We point out that the body can be made by two or more disconnected part, provided that the material composition is the same. We denote the total electric field as ${\bf E} \left( {\bf r} \right)$. The scattered electric ${\bf E}_S \left( {\bf r} \right)$, which is  defined in the whole space as ${\bf E}_S =  {\bf E} - {\bf E}_i $, is expressed as \cite{Forestiere16}
\begin{equation}
{\bf E}_S \left( {\bf r} \right) = \left( \varepsilon_R - 1 \right)  \sum_{\alpha=1}^\infty \frac{ \mathcal{P}_\alpha \left\{ {\bf E}	_i \right\}}{\varepsilon_\alpha - \varepsilon_R} {\bf w}_\alpha \left( {\bf r} \right),
\label{eq:DimerModeExpansion}
\end{equation}
where
\begin{align}
 &  \mathcal{P}_\alpha \left\{ {\bf E}_i \right\} = \frac{ \langle  \eigw^*, {\bf E}_i \rangle_\Omega}{ \langle \eigw^* ,{\bf \eigw}\rangle_\Omega},\label{eq:Projection}\\
&\langle {\bf f}, {\bf g} \rangle_{\Omega} = \int_\Omega {\bf f}^* \cdot {\bf g} \,dV.
\label{eq:ScalarProduct}
\end{align}
The set of complex poles $\left\{ \eig \right\}$, denoted as {\it eigen-permittivities} of the body, and the set of complex field modes $\left\{ \eigw \right\}$ do not depend on the body permittivity $\varepsilon_R$, but only on the shape of the body and its normalized size parameter $x$.  In particular, the eigen-permittivity $\varepsilon_\alpha$ is the value of the permittivity that the body should have such that the corresponding electric field mode $\eigw \left( {\bf r} \right)$ is a source-free solution of the Maxwell’s equations satisfying the Silver-M\"uller conditions at infinity. The modes are not orthogonal in the usual sense, but  they are bi-orthogonal, i.e.
\begin{equation}
   \langle {\bf w}_\alpha^*, {\bf w}_\beta \rangle = 0 \quad \forall \alpha \ne \beta.
\end{equation}
The imaginary part of $\eig$ is proportional to the averaged flux toward the infinity of the Poynting vector associated to the mode $\eigw$: it is always negative due to the Silver-Muller condition at infinity. The real part of the eigen-permittivity $\eig$ has not a definite sign \cite{Forestiere16}. 
The coefficient $\mathcal{P}_\alpha \left\{ {\bf E}_i \right\}$ accounts for the coupling of the external excitation with the mode ${\bf w}_\alpha \left( {\bf r} \right) $. Expression \ref{eq:DimerModeExpansion} disentangles the geometric and the material properties of the body and effectively predicts the resonant behavior of 3D bodies as their shape, size and permittivity vary.  Since $\im{ \eig \left( \omega \right)}<0$, the quantity $\eig \left( \omega \right) - \varepsilon_R \left( \omega \right)$ does not vanish in passive materials (where $\im{ \varepsilon_R \left( \omega \right) > 0 }$) as $\omega$ varies.  Nevertheless, for fixed geometry of the body, the amplitude of the mode $\eigw \left( {\bf r}\right)$ reaches its maximum in a neighbourhood of the frequency $\omega_\alpha$ such that
\begin{equation}
\underset{\omega}{\mbox{min}} \left| \frac{\eig \left( \omega \right) - \varepsilon_R \left( \omega \right)}{\varepsilon_R \left( \omega \right) - 1} \right| = \rho_\alpha.
\label{eq:ResonantCondition}
\end{equation}
This is the resonant condition for the mode $\eigw \left( {\bf r } \right)$: $\omega_\alpha$ is the resonance frequency of the mode $\eigw \left( {\bf r } \right) $  and $\rho_\alpha$ is the corresponding residuum. They do not depend on the incident field. The width of the corresponding resonance is related to the value of the residuum. Specifically, a larger residuum is associated with a broader resonance. The coupling coefficient $\mathcal{P}_\alpha \left\{ {\bf E}_i \right\}$ also depends on the frequency but it varies very slowly if compared with $1/\left( \varepsilon_\alpha - \varepsilon_R \right)$. The modes with $\re{\eig} > 0 $ can be  resonantly excited in dielectrics, while the modes with $\re{\eig} < 0 $ can be  resonantly excited in metals.

The modes are solenoidal in the body region $\Omega$ and in the exterior region, but their normal component to the body surface is discontinuous unless it is equal to zero. They can be classified according to their behaviour in the limit $x \rightarrow 0$. There is a set of modes that for $x \rightarrow 0$ become irrotational everywhere and have discontinuous normal component to the body surface (quasi-stationary electric modes or plasmonic modes). In this paper, we call them longitudinal modes; we denote them with $\left\{ \eigw^\parallel \left( {\bf r } \right) \right\}$ and the corresponding eigen-permittivity with $\left\{ \varepsilon^\parallel_\alpha \right\}$ . In addition to these modes, there is another subset of modes that for $x \rightarrow 0 $ have normal component to the body surface equal to zero. In this limit, these modes become solenoidal everywhere, but they are not irrotational (quasi-stationary magnetic modes).
In this paper, we call them transverse modes; we denote them with $\left\{ \eigw^\perp \left({\bf r} \right) \right\}$ and the corresponding eigen-permittivity with $\left\{ \varepsilon^\perp_\alpha \right\}$. For $x \rightarrow 0 $ we found that $\re{\varepsilon^\parallel_\alpha} < 0$ and  $\re{\varepsilon^\perp_\alpha} > 0$. {For finite $x$, we have that the eigenvalues ${\varepsilon^\parallel_\alpha \left( x \right)}$ move within a finite region of the complex plane, while  $\re{\varepsilon^\perp_\alpha}$ always remains positive.}
The plasmon hybridization theory only considers the longitudinal modes. The transverse modes play a fundamental role in the scattering from dielectric structures.

\subsection*{Isolated sphere}
Now, we briefly resume the properties of the material independent modes and the  corresponding eigen-permittivities of an isolated-sphere with radius $R$, and size parameter $x = 2\pi R/\lambda$. They can be expressed analytically in terms of the vector spherical wave functions (VSWF) \cite{Forestiere16}. They depend on five indexes $\left\{ \delta, p, m, n, l \right\}$ and we indicate them with ${\bf e}_{pmnl}^\delta \left( {\bf r}\right)$. The superscript $\delta$ distinguishes between the transverse magnetic (TM) modes (electric type modes) and transverse electric (TE) modes (magnetic type modes). The subscript $p$ distinguishes between even ($e$) and odd ($o$) modes with respect to the azimuthal variable. The numbers $n\in \mathbb{N}$ and $0\le m \le n$  characterize the angular dependence of the modes: $m$ is the number of oscillations along the azimuth and $n$ is the multipolar order. The mode number $l \in \mathbb{N}$ gives the number of maxima of the mode amplitude along the radial direction inside the sphere. Due to the spherical symmetry, the eigen-permittivities only depend on the indexes $\left\{ n,l \right\}$. In the body region ${\bf e}_{pmnl}^\mathtt{TM} \left( {\bf r} \right) = {\bf N}_{pmn}^{(1)} \left( \sqrt{\varepsilon_{nl}^\mathtt{TM}} k_0 {\bf r} \right)$, ${\bf e}_{pmnl}^\mathtt{TE} \left( {\bf r} \right) = {\bf M}_{pmn}^{(1)} \left( \sqrt{\varepsilon_{nl}^\mathtt{TE}} k_0 {\bf r}  \right)$ where $ \varepsilon_{pmnl}^\mathtt{TE}$,  $\varepsilon_{pmnl}^\mathtt{TM} $ are the corresponding eigen-permittivities, $\NN{pmn}{1}$ and $\MM{pmn}{1}$ are the VSWF regular at the center of the sphere. 
The modes with $n=1$ are the dipolar modes, those with $n=2$ are the quadrupolar modes, and so on. We denote the electric and magnetic type modes as fundamental when $l=1$, and as higher order modes when $l>1$. 
It is worth noting that higher order electric modes and magnetic modes are not contemplated by the quasi-electrostatic resonance theory \cite{Mayergoyz05}, and it is not possible to include them within the quasi- electrostatic framework by simple using perturbation techniques. The eigen-permittivities $\varepsilon_{nl}^\mathtt{TM}$ and $\varepsilon_{nl}^\mathtt{TE}$ are the roots of two power series, which are given analytically in \cite{Forestiere16}. In particular, it results that $\left.\re{\varepsilon_{nl}^\mathtt{TM}}\right|_{l\ne 1}>0$  and $\left.\re{\varepsilon_{nl}^\mathtt{TE}}\right|>0$, while in the limit $x \rightarrow 0 $ it results that $\re{\varepsilon^\mathtt{TM}_{n,l=1}}<0$. 
The set of longitudinal modes $\left\{ {\bf w}_\alpha^\parallel \left( {\bf r} \right) \right\}$ coincides with the set of fundamental electric type modes $\left\{ {\bf e}_{pmn1}^\texttt{TM}  \right\}$ and the set of transverse modes $\left\{ {\bf w}_\alpha^\perp \left( {\bf r} \right) \right\}$ coincides with the set $\left\{ \left. {\bf e}_{pmnl}^\texttt{TM} \right|_{l>1} , {\bf e}_{pmnl}^\texttt{TE} \right\}$.

\begin{figure*}
\centering
	\includegraphics[width=\textwidth]{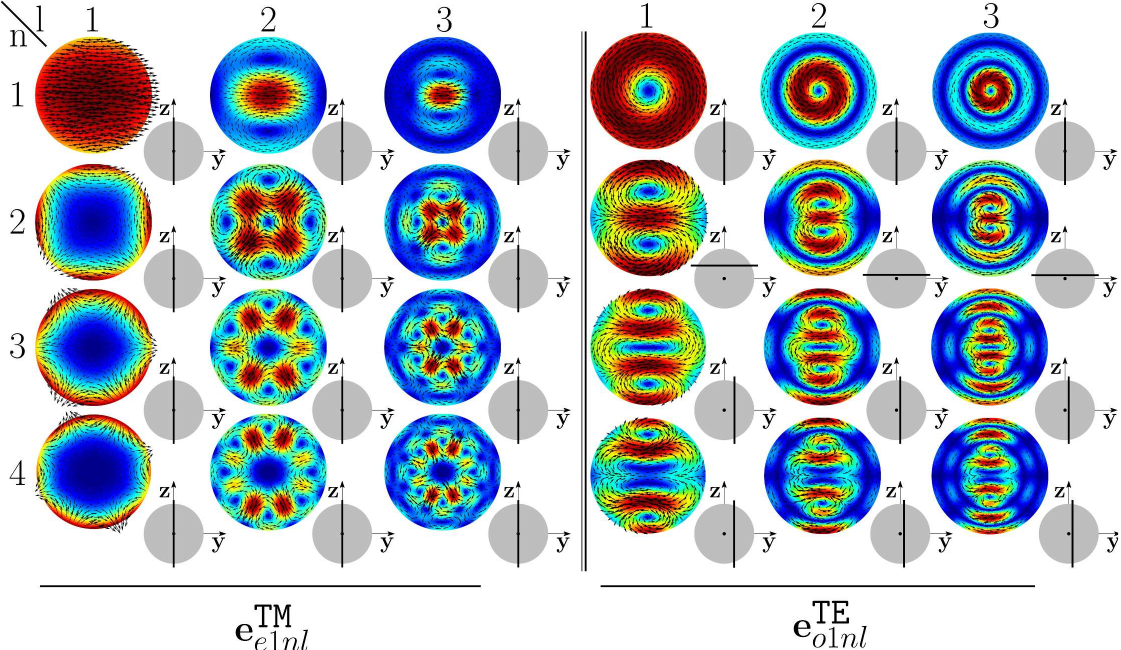}
\caption{Projections of the TM modes $\Ca{e1nl}$  and of the TE modes $\Cb{o1nl}$ of a sphere of radius $R = \lambda/4$ on the section shown in the bottom-right inset. The modes ${\bf e}_{p1n1}^\mathtt{TM}$ belong to the set $\left\{ {\bf w}_\alpha^\parallel \right\}$, while the remaining ones belong to the set $\left\{ {\bf w}_\alpha^\perp \right\}$.}
\label{fig:SingleModes}
\end{figure*}

\begin{figure*}
\centering
\includegraphics[width=0.75\textwidth]{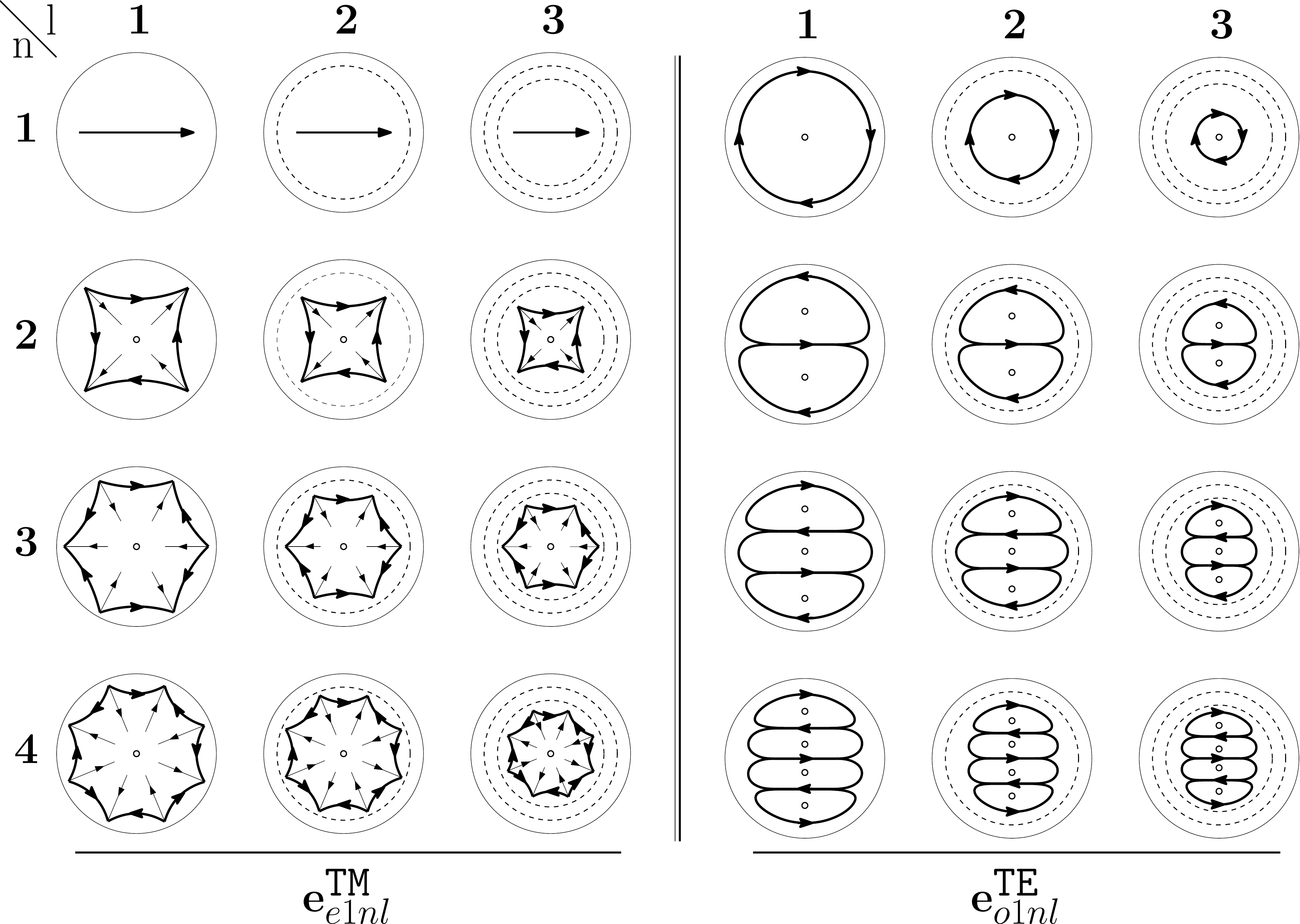}
\caption{Stylized version of the TM and TE modes of an isolated-sphere.}
\label{fig:StylizedSingle}
\end{figure*}

As an example, in Fig. \ref{fig:SingleModes} we show the TM modes $\Ca{e1nl}$, and TE  modes $\Cb{o1nl}$, of a sphere with radius $R = \lambda/4$, and corresponding size parameter $x=\pi/2$, for $n=1,2,3,4$ and $l=1,2,3$. Specifically, we plot the  real part of the mode components which are tangent to the section of the sphere shown in the inset on the bottom-right of each mode.  These modes are the building blocks of the dimer-modes, and we will make an extensive use of them in the following sections. Since these modes are weakly sensitive to changes of the size parameter $x$ we introduce a stylized version of them in Fig. \ref{fig:StylizedSingle}.
The TM mode $\Ca{e111}$, shown in the top-left corner of Fig. \ref{fig:SingleModes}, exhibits a dipolar character. Increasing the index $l$, while keeping fixed the order $n=1$, we observe two and three oscillations of the mode along the (vertical) radial direction for $l=2$ and $l=3$, respectively. Moreover, as $l$ increases, the region where the field is localized is increasingly squeezed in proximity of the sphere center. For these reasons, the stylized representation of the $\Ca{pm1l}$ is a single arrow, representing the electric dipole, enclosed by $l$ circles, and squeezed as $l$ increases.
The mode $\Ca{e121}$ shows a quadrupolar character with two sources and two sinks of the field lines, whereas the  mode $\Ca{e131}$ is of octupolar type with three sources and three sinks. In both cases, by increasing $l$ the number of oscillations along the radius increases. The stylized representations of $\Ca{pm2l}$ and $\Ca{pm3l}$ visually highlight both the number of sources-sinks, and the number of oscillations along the radial direction.
The mode $\Cb{o111}$ has a magnetic dipole character: it exhibits one vortex, associated to a magnetic dipole moment directed orthogonally to the vortex plane. By increasing $l$, we note that one ($l=2$) or two ($l=3$) additional contra-rotating vortices arise. For this reason, the stylized representation of the mode $\Cb{pm1l}$ is a current loop, enclosed by $l$ concentric circles. By increasing $n$ to 2, the mode $\Cb{o121}$ shows two identical vortices with antiparallel magnetic dipole moments. Also in this case, by increasing $l$ additional vortices appear. The number of oscillations of the mode along the radial (vertical) direction is $l$. The field lines of the magnetic octupole ($n=3,l=1$) and the magnetic hexadecapole ($n=4,l=1$) form three and four identical vortices, respectively. As we increase $l$, the number of oscillations of the mode along the radial (vertical) direction increases. The stylized representations of the TE modes highlight the number of vortices of the mode, and the number of oscillations along the radial direction. 
 By using Eq. \ref{eq:ResonantCondition} we find the resonant frequencies of the low order modes of both an Ag and a Si isolated-sphere (in Tabs. \ref{tab:ResonancesAgSingle} and \ref{tab:ResonancesSiSingle} of SI). Then, we untangle the scattering cross section from isolated Si and Ag spheres in terms of the contribution of different modes (Figs. \ref{fig:CscaAgSingle} and \ref{fig:CscaSiSingle} of the SI).

\subsection*{Sphere dimer}
\begin{figure}[htpb]
\centering
\includegraphics[scale=1]{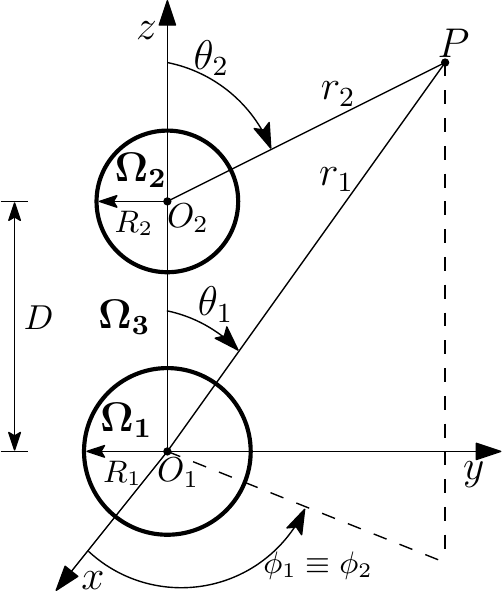}
\caption{Dimer composed by two spheres of radii $R_1$ and $R_2$ and center-center interparticle distance $D$.}
\label{fig:Domain_dimer}
\end{figure}

Now, let us consider the electromagnetic scattering from a dimer of spheres surrounded by vacuum. The spheres have radii $R_1$ and $R_2$, and a center-center separation $D$. They occupy the regions $\Omega_1$ and $\Omega_2$, while the surrounding space is denoted as $\Omega_3$, as shown in Fig. \ref{fig:Domain_dimer}. The two spheres are made of the same material. 
We define the dimensionless quantities
\begin{equation}
\begin{aligned}
x_1&= 2 \pi R_1 / \lambda,\\
x_2&= 2 \pi R_2 / \lambda,\\
d&= 2 \pi D / \lambda.\\
\end{aligned}
\end{equation}
As for the general case, the dimer-modes can be subdivided into two subsets: the subset $\left\{ {\bf d}_{\alpha}^\perp \left( {\bf r } \right) \right\}$ with eigen-permittivity $\left\{ \varepsilon^\perp_{\alpha} \right\}$ and the subset $\left\{ {\bf d}^\parallel_\alpha \left( {\bf r } \right) \right\}$ with eigen-permittivity $\left\{ \varepsilon^\parallel_\alpha \right\}$. They have the properties already stated for the general case. 
Now we determine the modes of the dimer by using as basis functions the modes of the isolated-spheres. 
We denote the modes of the $j$-th sphere, assumed to be isolated, as $\Cah{pmnl}$ and $\Cbh{pmnl}$ with $j \in \left\{ 1 , 2 \right\}$.
Then, we have (see Methods):
\begin{equation}
{\bf d}^{ \nu }_{pmq} ({\bf r}) = 
\displaystyle\sum_{n l} \CoeffdTM{pmq \, nl} \, \Cah{pmnl} \left( {\bf r} \right)
+\CoeffdTE{\bar pmq \, nl} \,
\Cbh{\bar pmnl} \left( {\bf r} \right)
\qquad \forall {\bf r} \in \Omega_j \quad \mbox{with} \quad j =1,2
\label{eq:DimerModeHybridization}
\end{equation}
where $\CoeffdTM{pmq \, nl}$ and $\CoeffdTE{pmq \, nl}$ are the projections of the dimer-mode on the conjugate of the $j$-th isolated-sphere mode
\begin{equation}
\begin{aligned}
 \CoeffdTM{pmq \, nl} & =  \mathcal{P}^{\mathtt{TM}|j}_{pmq \, nl} \left\{ \CC{pmq}({\bf r}) \right\} ,
  \\
  \CoeffdTE{pmq \, nl}& =  \mathcal{P}^{\mathtt{TE}|j}_{pmq \, nl} \left\{ \CC{pmq}({\bf r}) \right\} ,
 \end{aligned}
 \label{eq:HybridizationCoeff}
\end{equation}
where
\begin{equation}
\label{eq:ProjectionOnSingle}
\begin{aligned}
\Proj{TM}{\cdot} &= \frac{\langle \left( \Cah{pmnl} \right)^*, \cdot\,\, \rangle_{\Omega_j}}{\langle 
 \left( \Cah{pmnl} \right)^*, \Cah{pmnl} \rangle_{\Omega_j}}, \\
\Proj{TE}{\cdot} &= \frac{\langle 
 \left( \Cbh{pmnl} \right)^*, \cdot\,\, \rangle_{\Omega_j}}{\langle  \left( \Cbh{pmnl} \right)^*,  \Cbh{pmnl}  \rangle_{\Omega_j}}.
\end{aligned}
\end{equation}
The dimer-modes and the corresponding eigen-permittivities depend on four indices: $\nu \in \left\{ \parallel, \perp \right\}$, $ p \in \left\{ e, o \right\}$ ( in Eq. \ref{eq:DimerModeHybridization} the symbol $\bar{p}$ indicates the complement of $p$), $m \in \mathbb{N}_0$, and $q\in \mathbb{N}$.  
In the limit $ x \rightarrow 0$ the eigen-permittivities $\varepsilon^\nu_{pmq}$ are sorted in ascending order of $q$ for any given $\nu$, $p$ and $m$. The same order is kept for finite values of $x$, by following $\varepsilon^\nu_{pmq}$ on the complex plane as $x$ varies. 
In order to weight the contribution of the isolated-sphere modes $\Cah{pmnl}$ and $\Cbh{pmnl}$ of the $j$-th sphere in the expansion \ref{eq:DimerModeHybridization} of the dimer-mode $\CC{pmq}^\nu$, we introduce the following synthetic parameters: 
\begin{equation}
 \CoeffhTM{pmq \, nl} =   \underset{{\bf r} \in \Omega_1 \cup \Omega_2}{ \max} \left\| \re{\CoeffdTM{ pmq \, nl} \Cah{pmnl} \left( {\bf r} \right) } \right\|
 \label{eq:HybridazionCoeffTM}
\end{equation}

\begin{equation}
\CoeffhTE{\overline pmq \, nl} = \underset{{\bf r} \in \Omega_1 \cup \Omega_2}{ \max} \left\| \re{\CoeffdTE{\overline pmq \, nl} \Cbh{\overline pmnl} \left( {\bf r} \right) } \right\|
  \label{eq:HybridazionCoeffTE}
\end{equation}
The parameter $\CoeffhTM{pmq \, nl}$ ($\CoeffhTE{\overline pmq \, nl}$) represents the maximum magnitude of the real part of the dimer-mode  within the $j$-th sphere (which corresponds to the amplitude at $t=0$) that we would have if only the mode $\Cah{pmnl}$ ($\Cbh{\overline pmnl}$) in the expansion \ref{eq:DimerModeHybridization} were considered. In the following sections, for any given dimer-mode $\CC{pmq}^\nu$  we normalize the parameter $\CoeffhTM{pmq \, nl}$ and $\CoeffhTE{\overline pmq \, nl}$ to the overall maximum, i.e. $ \underset{nl}{\max}\left\{ \CoeffhTM{pmq \, nl}, \CoeffhTE{\overline pmq \, nl}  \right\}$. These normalized parameters will be denoted as 
$\CoeffhhTM{pmq \, nl}$ and $\CoeffhhTE{\overline pmq \, nl}$. We call them hybridization weights. Moreover, if $R_1=R_2$, for symmetry considerations we also have that  $\CoeffhhhhTM{pmq \, nl} = \CoeffhhhTM{pmq \, nl}{1} = \CoeffhhhTM{\overline pmq \, nl}{2}$ and $\CoeffhhhhTE{pmq \, nl} = \CoeffhhhTE{pmq \, nl}{1} = \CoeffhhhTE{\overline pmq \, nl}{2}$ 

%% file: AgZ_R4/AgZ.tex
\subsection*{Longitudinally polarized Ag homo-dimer}
\begin{figure*}[h!!]
\centering
\includegraphics[width=0.9\textwidth]{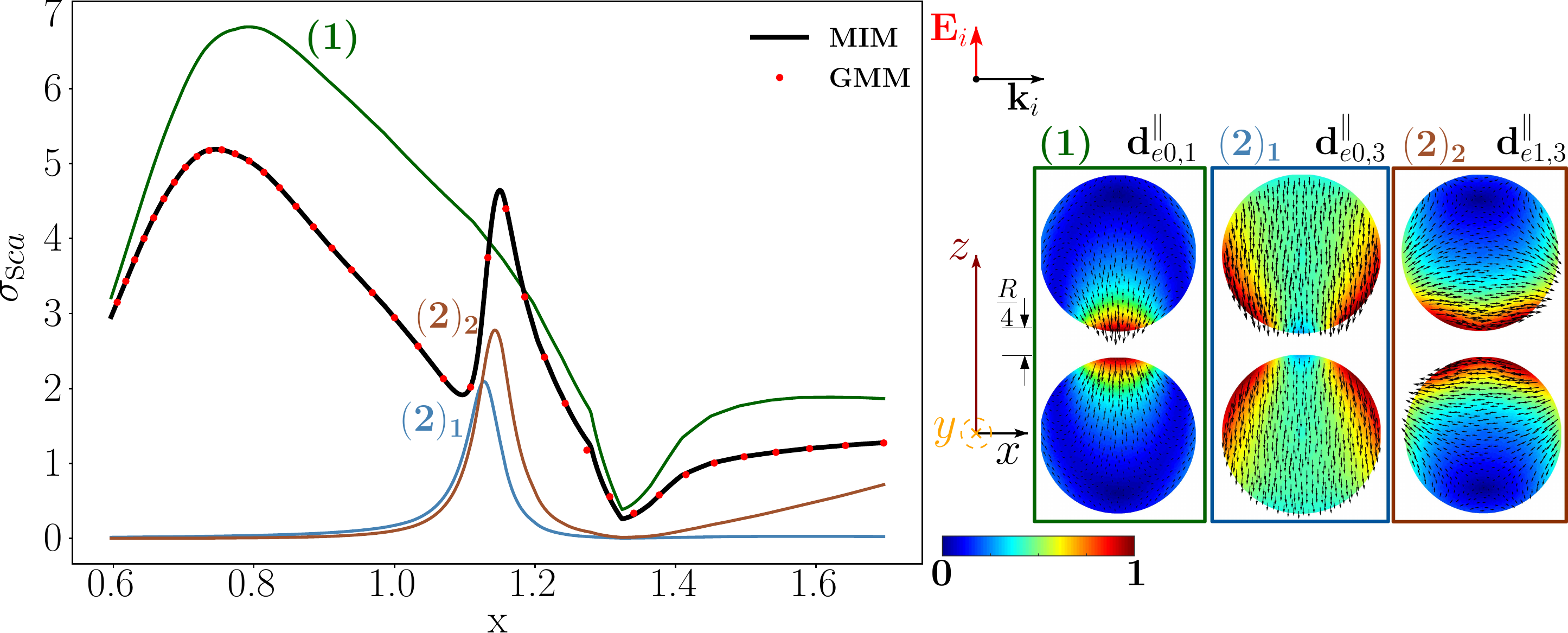}
\caption{Scattering efficiency $\sigma_{sca}$ of an Ag homo-dimer as a function of the spheres size parameter $x=2\pi R/\lambda$, obtained via material-independent-mode expansion (black line) and by direct-calculation (red dots). The radius of each sphere is $R=67.5$ nm, the edge-edge distance $R/4=16.9$ nm. The dimer is excited by a plane wave polarized along the dimer axis $\hat{\bf z}$ and propagating along the transverse direction $\hat{\bf x}$. Partial scattering efficiencies (in color) of the three dominant dimer-modes whose $y=0$-plane projections are shown on the right.}
\label{fig:SilverCscaZ}
\end{figure*}

\begin{table*}[h!]
\centering
\begin{tabular}{|c||ccccc||cc|}
\hline
mode  & $x_{emq}^\parallel$ & $\omega_{emq}$ (Prad/s) & $ {\varepsilon}^\parallel_{emq}$  &  $\varepsilon_{Ag}$ & $\rho^\parallel_{emq}$ & \# peak & x \\
\hline
$\CCo{e0,1}$  & 0.650 & 2.89 & -4.16 -3.46i & -19.62 +0.46i & 0.773 & $(1)$ & 0.747
 \\
$\CCo{e0,3}$  & 1.126 & 5.00 & -3.02-0.42i & -3.21 +0.20i & 0.155 & $(2)$ & 1.15
 \\
$\CCo{e1,3}$  & 1.139 & 5.06 & -2.79 -0.40i & -2.97 + 0.22i& 0.163 & $(2)$ & 1.15
 \\
\hline
\end{tabular}
\caption{Resonant size parameter $x$, corresponding values of the resonant frequency, eigen-permittivity, Ag permittivity, and residuum of the dimer-modes which dominate the scattering efficiency of Fig. \ref{fig:SilverCscaZ}. Positions of the peaks of the total scattering efficiency.}
\label{tab:SilverZ}
\end{table*}
First, we study a silver homo-dimer with $R=67.5nm$, and edge-edge separation $R/4 = 16.875nm$. We consider the modes that are excited by an incident field that is polarized along the dimer axis $\hat{\bf z}$, while it is propagating along the transverse direction $\hat{\bf x}$. In Fig. \ref{fig:SilverCscaZ}, we plot the scattering efficiency $\sigma_{sca}$ obtained by using the material-independent-mode (MIM) expansion \ref{eq:DimerModeExpansion} (black line), and by a direct calculation (red dots) as a function of the size parameter $x$. For the direct calculation we use the code ``Generalized Multiparticle Mie-Solution (GMM)" by Yu-lin Xu \cite{xu95}. Specifically, we use in the GMM two VSWF sets centred in the two spheres each of them described by Eq. (4) of Ref. \cite{xu95} with $ 1 < n \le 8$. The two results are in very good agreement. As expected, the scattering efficiency is significantly different from the one obtained in the quasi-electrostatic limit (Q-ES) approximation \cite{Mayergoyz05} which is shown in the SI. We also show in color the partial scattering efficiencies of three dominant dimer-modes, whose projections (real part) on the $y=0$ plane are represented on the right. The partial scattering efficiency is the scattering efficiency that we would have if only one dimer-mode is excited at a time.  It is important to note that the total scattering efficiency is not the sum of the partial scattering efficiencies because the modes with same index $m$ may interfere. Nevertheless, the partial scattering efficiencies enable us to identify the dimer-modes responsible for each peak of the total scattering efficiency. We list in Tab. \ref{tab:SilverZ} the resonant size parameters and corresponding resonant frequencies of the three dominant modes, which are evaluated by Eq. \ref{eq:ResonantCondition}. It is worth noting that these values do not depend on the chosen excitation. The table also highlights that the residuum associated to the dimer-mode $\CCo{e0,1}$, which causes the first peak, is larger than the corresponding residuum of $\CCo{e0,3}$ and $\CCo{e1,3}$ associated to peaks $(2)_1$ and $(2)_2$, thus the corresponding resonance is broader, as confirmed by Fig. \ref{fig:SilverCscaZ}. We note that the relevant dimer-modes are all longitudinal modes. This is because, as stated in the previous section,  the numerator of Eq. \ref{eq:ResonantCondition} is always very large since the real part of the eigen-permittivities of the transverse dimer-modes is always positive, while $\re{\varepsilon_{R,Ag}}<0$ in the visible.

In the SI we analyse the scattered electric field within the two spheres calculated at the two peaks of the scattering efficiency (Fig. \ref{fig:SilverZstate}). The corresponding near field profiles resemble the modes $\CCo{e0,1}$ and $\CCo{e1,3}$, respectively, which indeed dominate the scattering in the near zone as well.

%
\begin{figure*}[h!]
\centering
\includegraphics[width=1\textwidth]{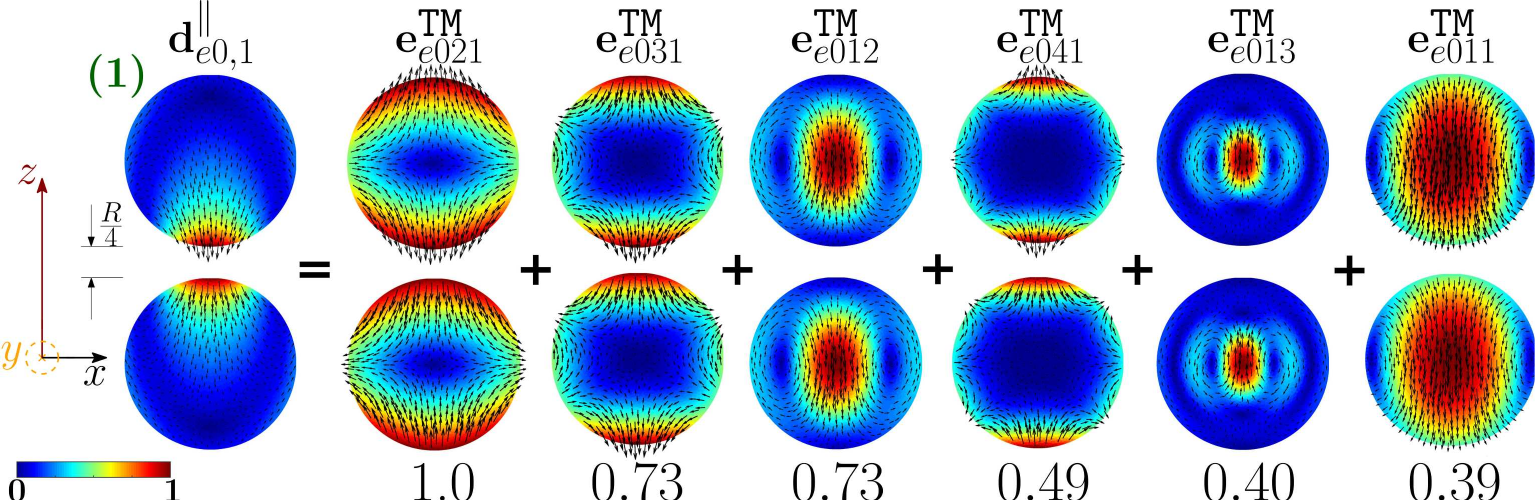}
\caption{Decomposition of the dimer-mode $\CCo{e0,1}$  at $x=0.747$, in terms of hybridizing isolated-sphere modes (real part of the projection on the $y=0$ plane). Each isolated-sphere mode is multiplied by the corresponding expansion coefficient of Eq. \ref{eq:HybridizationCoeff}. Below each isolated-sphere mode its hybridization weight $\CoefflTM{e01\, nl}$ is shown.}
\label{fig:SilverZPeak1}
\end{figure*}

The analysis of the partial scattering efficiencies  reveals that the dimer-mode $\CCo{e0,1}$ dominates the total scattering efficiency at its first peak.
The mode $\CCo{e0,1}$ originates from the hybridization of the isolated-sphere modes shown in Fig. \ref{fig:SilverZPeak1}, which include both longitudinal and transverse modes. The figure also gives the corresponding hybridization weights $\CoefflTM{e01\, nl}$. Specifically, the fundamental electric quadrupole $\Ca{e021}$, octupole $\Ca{e031}$, hexadecapole $\Ca{e041}$, and dipole $\Ca{e011}$ interfere constructively in the proximity of the dimer gap. The same modes interfere destructively in the regions which are located diametrically opposite to the gap. In the central region of each sphere, the fundamental $\Ca{e011}$, second order $\Ca{e012}$, and third order $\Ca{e013}$ electric dipoles also interfere destructively. %
It is worth noting that, among the isolated-sphere modes that take part in the hybridization process, only the transverse modes $\Ca{e012}$ and $\Ca{e013}$ are not included within the set of isolated-sphere modes predicted by the Q-ES approximation \cite{Mayergoyz05}. However, they give a negligible contribution to the far field. Indeed, the Q-ES approximation is also able to predict that the mode $\CCo{e0,1}$ causes the first $\sigma_{sca}$ peak, as demonstrated in the SI.  
%
%
%
\begin{figure*}[h!]	
\centering
	\includegraphics[width=0.6\textwidth]{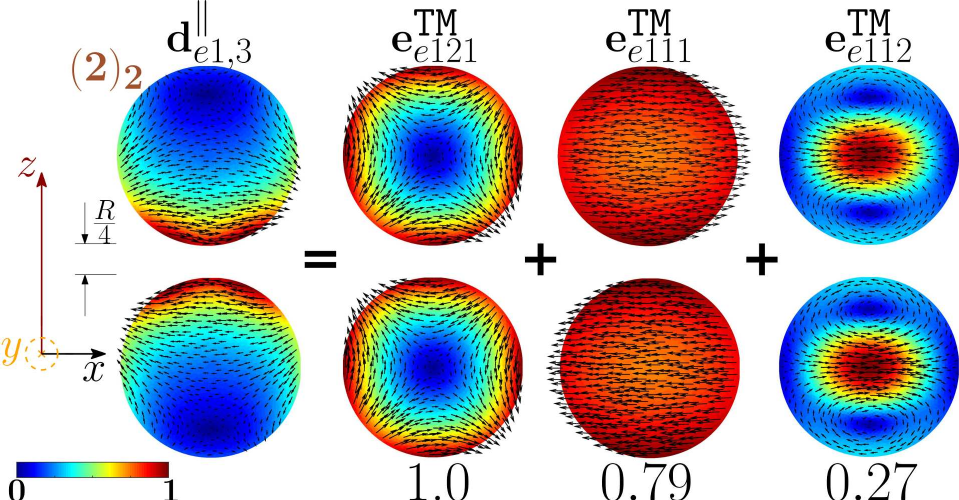}
\caption{Decomposition of the dimer-mode $\CCo{e1,3}$ at $x=1.15$  in terms of hybridizing isolated-sphere modes (real part of the projection on the $y=0$ plane). Each isolated-sphere modes is multiplied by the expansion coefficients of Eq. \ref{eq:HybridizationCoeff}. Below each isolated-sphere mode we also show its hybridization weight $\CoefflTM{e13\, nl}$.
}
\label{fig:SilverZPeak2}
\end{figure*}

We now examine the second peak of the scattering efficiency. From the analysis of the partial scattering efficiencies, and of the resonances shown in Tab. \ref{tab:SilverZ}, we conclude that it is dominated by two different modes, i.e. $\CCo{e0,3}$ and $\CCo{e1,3}$. Their resonance positions $x_{e0,3}^\parallel = 1.126$ and $x_{e1,3}^\parallel = 1.139$ are both located in proximity of the second $\sigma_{sca}$ peak at $x=1.15$. In particular, the mode $\CCo{e1,3}$ provides the largest  contribution. We analyse in Fig. \ref{fig:SilverZPeak2} its decomposition in terms of the hybridizing isolated-sphere modes, providing the corresponding hybridization weights $\CoefflTM{e13\, nl}$. Also in this case both longitudinal and transverse isolated-sphere modes contribute to the hybridization.
The fundamental electric quadrupole $\Ca{e121}$, and dipole $\Ca{e111}$ interfere constructively in the proximity of the gap, while they undergo destructive interference in the regions of the two spheres diametrically opposite to the gap. In the central region of both spheres, we observe destructive interference between the fundamental $\Ca{e111}$, and second order $\Ca{e112}$ electric dipole. 
From the analysis carried out in the SI, we note that the mode $\CCo{e1,3}$ is not excited under the Q-ES approximation because it has a vanishing total dipole moment.

As we vary the distance between two spheres, the dimer-mode  change. The advantage of the hybridization approach is now clear: the dimer-modes are represented in terms of elementary building blocks, i.e. the isolated-sphere modes, that do not change as the arrangement of the sphere changes; only the hybridization weights change. For instance, as the edge-edge distance is increased to $R$, the scattering efficiency varies (Fig. \ref{fig:SilverCscaZ_R} in SI). The dimer-mode that causes its first $\sigma_{sca}$ peak arises mainly from the fundamental electric dipole $\Ca{e011}$ (Fig. \ref{fig:SilverZPeak1_R} in SI), while the electric quadrupole $\Ca{e021}$ that was dominant for a gap size of $R/4$ now plays a minor role. On the contrary, the dimer-mode responsible for the second peak does not significantly change with respect to Fig. \ref{fig:SilverZPeak2} and it is still dominated by the fundamental electric dipole $\Ca{e111}$ and quadrupole $\Ca{e121}$ (Fig. \ref{fig:SilverZPeak2_R} in SI). For very large distances $D \rightarrow \infty$, the scattering efficiency of the dimer approaches the scattering efficiency of the isolated-sphere (Fig. \ref{fig:CscaAgSingle} in SI). This $\sigma_{sca}$ has two peaks, due to the fundamental electric dipole and quadrupole, respectively.

%

%% file: AgX_R4/AgX.tex
\subsection*{Transversely polarized Ag homo-dimer}

\begin{figure*}[h!]
\centering
\includegraphics[width=0.9\textwidth]{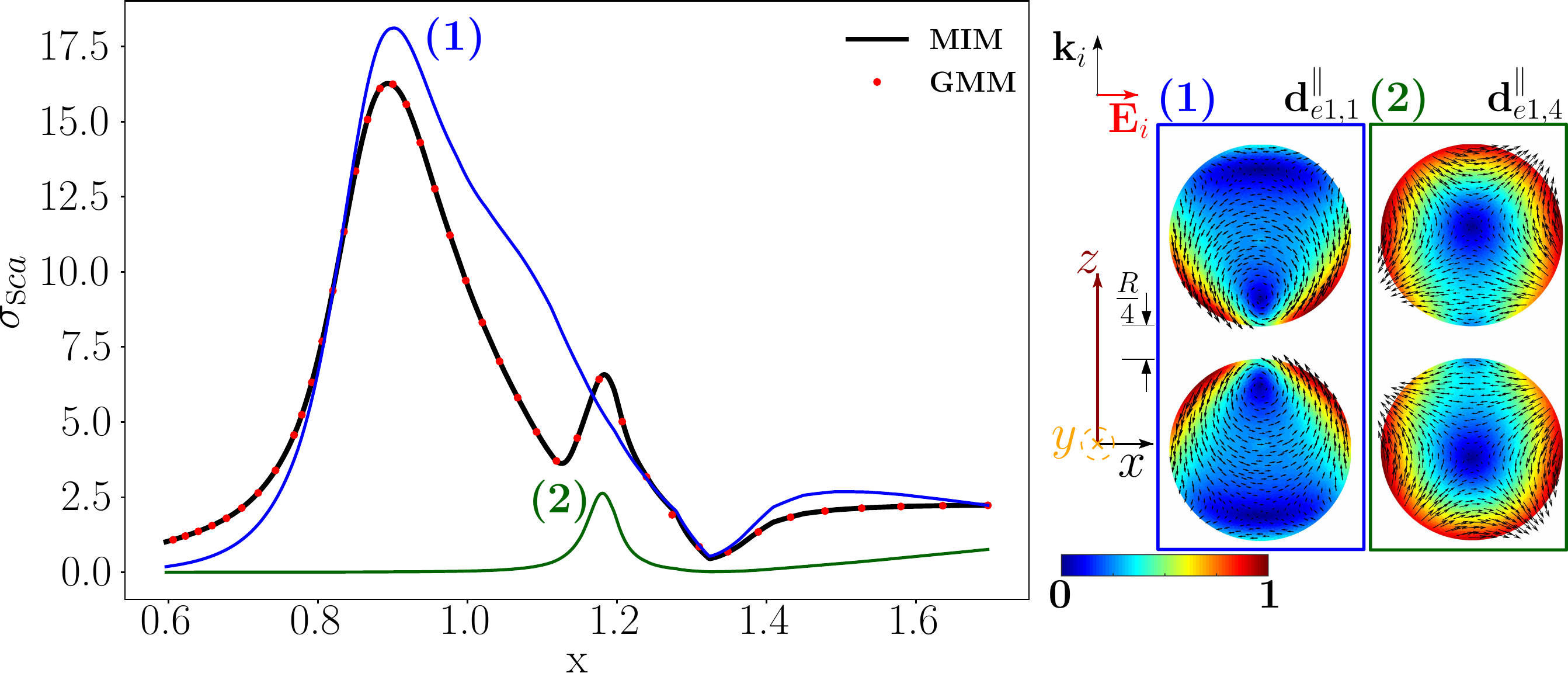}
\caption{Scattering efficiency $\sigma_{sca}$ of an Ag homo-dimer as a function of the size parameter $x = 2 \pi R/ \lambda$, obtained via material-independent-mode expansion (black line) and by direct-calculation (red dots). The radius of each sphere is $R=67.5$ nm, the edge-edge distance $R/4=16.9$ nm. The dimer is excited by a plane wave propagating along the dimer's axis and polarized along the transverse direction. Partial scattering efficiency (in color) of two dominant dimer-modes whose $y=0$ plane projections are shown on the right.}
\label{fig:SilverCscaX}
\end{figure*}

\begin{table*}[h!]
\centering
\begin{tabular}{|c||ccccc||cc|}
\hline
mode & peak  $x_{pme}^\parallel$ & $\omega_{pme}$ [Prad/s]&	 $\varepsilon^\parallel_{pme}$ & $\varepsilon_{r,Ag}$ & $\rho_{pme}^\parallel$ & \# peak & $x$ \\
\hline
$\CCo{e1,1}$ &  0.846 & 3.76 & -5.35-2.80i & -9.88 + 0.31i & 0.505 & (1) & 0.892
 \\
$\CCo{e1,4}$ & 1.176 & 5.22 & -2.18-0.22i & -2.33 + 0.26i & 0.152 & (2)  & 1.183
 \\
\hline
\end{tabular}
\caption{
Resonant size parameter $x$, corresponding value of the resonant frequency, eigen-permittivity, Ag permittivity, and residuum of the dimer-modes which dominate the scattering efficiency of Fig. \ref{fig:SilverCscaX}. Positions of the peaks of the total scattering efficiency.} 
\label{tab:SilverX}
\end{table*}

\begin{figure*}[h!]
\centering
\includegraphics[width=\textwidth]{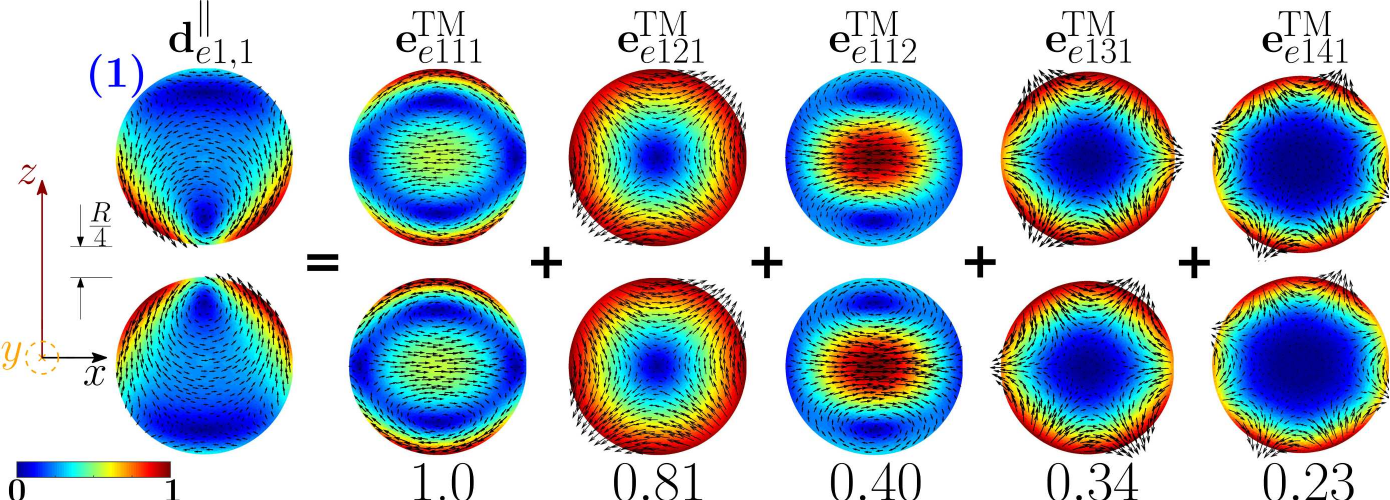}
\caption{Decomposition of the  dimer-mode
$\CCo{e1,1}$ at $x=0.892$ in terms of hybridizing isolated-sphere modes (real part of the projection on the $y=0$ plane). Each isolated-sphere modes is multiplied by the expansion coefficients of Eq. \ref{eq:HybridizationCoeff}. Below each isolated-sphere mode we also show its hybridization weight $\CoefflTM{e11\, nl}$.}
\label{fig:P1_Ag_X}
\end{figure*}

We now study an identical Ag homo-dimer, but we consider the modes that are excited by a plane-wave polarized along the $\hat{\bf x}$-direction, and propagating along the direction $\hat{\bf z}$. 
In Fig. \ref{fig:P1_Ag_X}, we plot the corresponding scattering efficiency obtained from both the mode expansion \ref{eq:DimerModeExpansion} (black line) and by the direct GMM calculation  (red dots). As in the previous case, the scattering efficiency is significantly different from the one obtained in the Q-ES limit \cite{Mayergoyz05}  (Fig. \ref{fig:EQS_Ag_X} in SI). We also show in color the partial scattering efficiency of the dimer-modes $\CCo{e1,1}$, $\CCo{e1,4}$ dominating the scattering response. Their projections on the $y=0$ plane (real part) are shown on the right. 
We list in Tab. \ref{tab:SilverX} the values of the resonant size parameter of the two dominant modes, together with the corresponding resonant frequency, eigen-permittivity, Ag permittivity, and residuum. It is apparent that the two resonant frequencies are in very close proximity to the two $\sigma_{sca}$ peaks, namely $x=0.892$ and $x=1.183$. The mode $\CCo{e1,1}$ exhibits a higher residuum than the mode $\CCo{e1,4}$: this is consistent with the fact that the corresponding partial scattering efficiency has a broader peak. {As in the longitudinally polarized Ag dimer (Fig. \ref{fig:SilverCscaZ}), the relevant dimer-modes are all longitudinal.}

In the SI we also plot within the two spheres the scattered electric field calculated at the two $\sigma_{sca}$ peaks (Fig. \ref{fig:SilverXstate}). The effects of the field propagations along the dimer axis are now important.  

The dimer-mode $\CCo{e1,1}$, responsible for the first $\sigma_{sca}$ peak, originates from the hybridization of the isolated-sphere modes shown in Fig. \ref{fig:P1_Ag_X}. In particular, the fundamental electric dipole $\Ca{e111}$ and quadrupole $\Ca{e121}$ interfere destructively in the close proximity of the gap. On the contrary, in the regions on the left and on the right of the dimer gap, the fundamental  quadrupole $\Ca{e121}$, octupole $\Ca{e131}$, hexadecapole $\Ca{e141}$ interfere constructively, determining a maximum in intensity. 

The mode $\CCo{e1,1}$ has a zero total dipole moment and it cannot be excited by a plane-wave in the Q-ES approximation. Nonetheless, in the presented full-Maxwell scenario its coupling to the plane-wave is different from zero. This is because the center-center distance between the two spheres is approximately one-third of the wavelength: thus the incident wavelength undergoes a phase inversion during its propagation within the dimer. The Q-ES approximation also fails to predict the scattering efficiency (Fig. \ref{fig:EQS_Ag_X} in SI): the Q-ES $\sigma_{sca}$ features one peak which is due to the mode $\CCo{e1,2}$ characterized by field lines all oriented in the same direction.

\begin{figure}[h!]
\centering
	\includegraphics[width=0.8\textwidth]{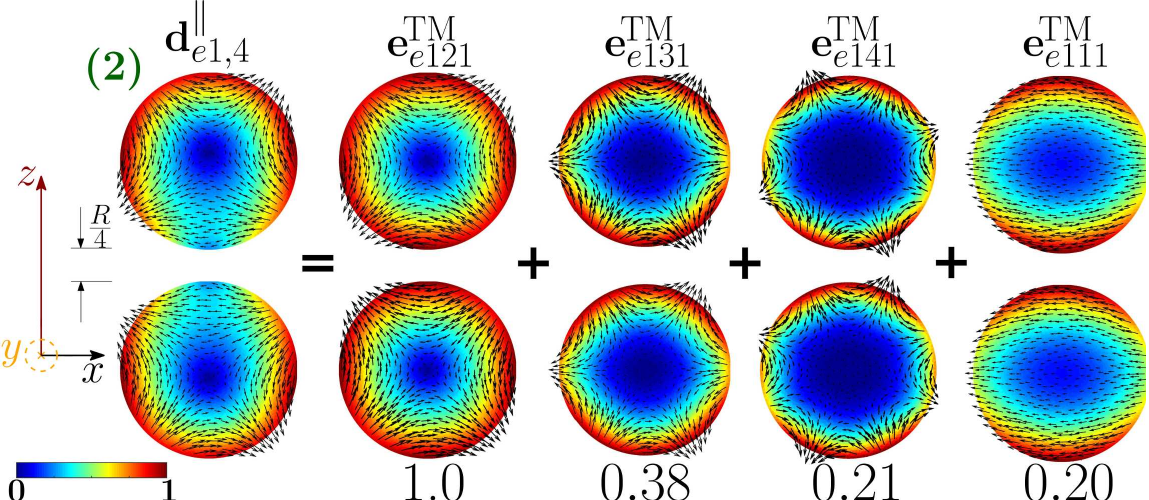}
\caption{Decomposition of the  dimer-mode
$\CCo{e1,4}$ at $x=1.183$, in terms of hybridizing isolated-sphere modes (real part of the projection on the $y=0$ plane). Each isolated-sphere modes is multiplied by the expansion coefficients of Eq. \ref{eq:HybridizationCoeff}. Below each isolated-sphere mode we also show its hybridization weight $\CoefflTM{e14\, nl}$.}
\label{fig:P2_Ag_X}
\end{figure}

The dimer-mode $\CCo{e1,4}$, which is behind the second $\sigma_{sca}$ peak, arises from the interaction among the isolated-sphere modes shown in Fig. \ref{fig:P2_Ag_X}. The fundamental electric quadrupole $\Ca{e121}$ and octupole $\Ca{e131}$ interfere destructively in the proximity of the gap and constructively within the region of each sphere opposite to the gap. A minor contribution to the hybridization process also comes from $\Ca{e141}$ and $\Ca{e111}$.

%

As the edge-edge distance between the two spheres increases to $R$ the scattering efficiency changes (Fig. \ref{fig:SilverCscaZ_R} in SI). The first $\sigma_{sca}$ peak is even more dominated by the fundamental electric dipole (Fig.   
 \ref{fig:SilverXPeak_R} in SI). The mode responsible for the second peak remains the fundamental electric quadrupole (Fig.   
 \ref{fig:SilverXPeak2_R} in SI). For very large distances $D \rightarrow \infty$, these two modes are the only ones to survive.

%% file: SiZ_R4/SiZ.tex
\subsection*{Longitudinally polarized Si homo-dimer}
\begin{figure*}[h!]
\centering
\includegraphics[width=\textwidth]{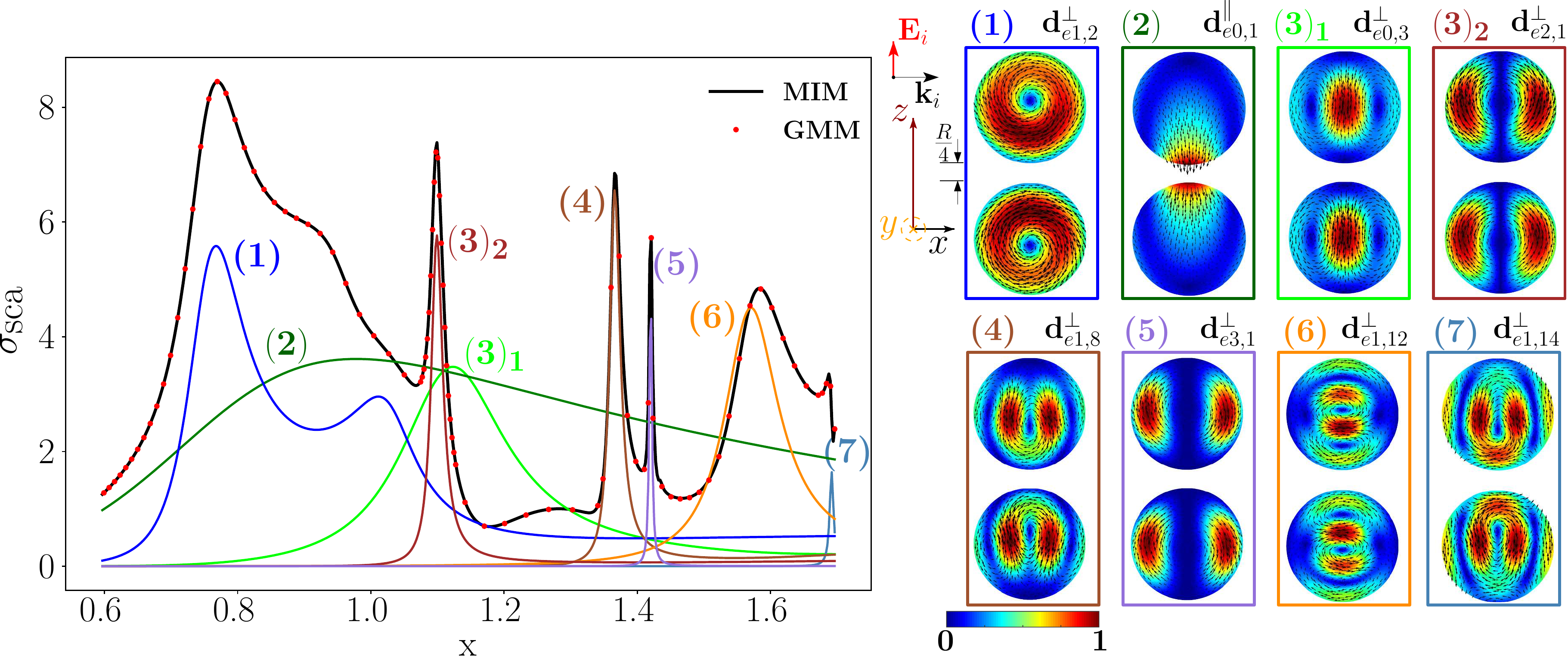}
\caption{Scattering efficiency $\sigma_{sca}$ of a Si-spheres homodimer as a function of the size parameter $x = 2 \pi R/ \lambda$, obtained via material-independent-mode expansion (black line) and by direct-calculation (red dots). The radius of each sphere is $R$, the edge-edge distance $R/4$. The dimer is excited by a plane wave polarized along the dimer's axis $\hat{\bf z}$ and propagating along the transverse direction $\hat{\bf x}$. Partial scattering efficiency (in color) of eight dominant dimer-modes whose $xz$-plane projections (real part) are shown on the right.}
\label{fig:Csca_Si_Z}
\end{figure*}

\begin{table*}[h!]
\centering
\begin{tabular}{|c||cccc||cc|}
\hline
mode  & $x_{emq}^\perp$ & $\omega_{emq}^\perp$ [Prad/s] & $\varepsilon_{emq}^\perp$ & $\rho_{emq}^\perp$ & \# peak & x \\
\hline
$\CCp{e1,2}$  & 0.760 & 2.29 & 16.11-2.23i & 0.149 &(1) & 0.771
 \\
$\CCo{e0,1}$  & 1.696 & 5.08 & -0.70 - 3.96i & 1.038 &(2) & 0.901
 \\
$\CCp{e0,3}$  & 1.126 & 3.37 & 15.67-2.50i & 0.168 &$(3)$ & 1.099
 \\
$\CCp{e2,1}$  & 1.099 & 3.29 & 15.99-0.32i & 0.021 &$(3)$ & 1.099
 \\
$\CCp{e1,8}$ & 1.366 & 4.09 & 16.00-0.26i & 0.017 & $ \left(4\right)$ & 1.336
 \\
$\CCp{e3,1}$ & 1.421 & 4.26 & 15.98-0.05i & 0.004 & $\left(5\right)$ &  1.423
 \\
$\CCp{e1,12}$  & 1.571 & 4.71 & 15.96-1.02i & 0.068 & $\left(6\right)$ & 1.584
 \\
$\CCp{e1,14}$  & 1.692 & 5.07 & 15.99-0.07i & 0.005 & $\left(7\right)$ & 1.688
 \\
\hline
\end{tabular}
\caption{Values of $x$ minimizing the residua, corresponding value of the resonant frequency (when $R=100nm$), eigenpermittivity, and residuum of the dimer-modes which dominate the scattering efficiency of Fig. \ref{fig:Csca_Si_Z}. Positions of the peaks of the total scattering efficiency.}
\label{tab:SiliconZ}
\end{table*}
Now, we study the scattering from a homo-dimer of the same geometry but made of a dielectric material, namely Silicon, with permittivity $\varepsilon_R=16$. We investigate the modes excited by a plane wave polarized along the dimer axis $\hat{\bf z}$, while it is propagating along the transverse direction $\hat{\bf x}$. In Fig. \ref{fig:Csca_Si_Z}, we plot the scattering efficiency obtained by the material-independent-mode expansion \ref{eq:DimerModeExpansion} (black line) and by the direct GMM calculation \cite{xu95} (red dots) as a function of the spheres size parameter $x$. It is worth to note that $\sigma_{sca}$ does not depend on $R$ and $\lambda$ separately, but only on $x$, because we have assumed that Si is not dispersive in time. We also show in color the partial scattering efficiencies of the eight dominant dimer-modes, whose real projections on the $y=0$ plane are represented on the right. 
It is worth to point out that by calculating $\sigma_{sca}$ using only these $8$ dimer-modes the agreement with the GMM remains satisfactory (Fig. \ref{fig:Csca_Si_Z_OPh} of SI).
 
We list in Tab. \ref{tab:SiliconZ} the resonant size parameter of the eight dominant modes and the corresponding resonant frequency (assuming $R=100nm$), eigen-permittivity and residuum.  We note that the residuum of the longitudinal mode $\CCo{e0,1}$ is almost one-order of magnitude higher than the residuum of any of the transverse modes in Tab. \ref{tab:SiliconZ}.  This fact can be explained by referring to Eq. \ref{eq:ResonantCondition}:
the real part of the eigen-permittivity $\varepsilon_{e0,1}^\parallel \left( x \right)$ never equates the Si eigen-permittivity since, irrespectively of $x$, we found that	 $-4.7 < \varepsilon_{e0,1}^\parallel \left( x \right) <0.49$ while $\varepsilon_{R,Si}=16$. On the contrary, for the transverse modes, there always exists  a value of $x$ in correspondence of which $\re{\varepsilon_{emq}^\perp \left( x \right)} = \varepsilon_{R,Si}$, as it is also apparent from the fourth column of Tab. \ref{tab:SiliconZ}. As a consequence, the partial scattering cross section of the mode  $\CCo{e0,1}$ is very broad compared to the ones of transverse modes.  
Nevertheless, the mode $\CCo{e0,1}$ has to be considered because it strongly couples with the plane wave and it also has  a stronger radiative strength \cite{forestiere2017nanoparticle}.

We note that only the transverse modes and the fundamental electric dipole of the isolated-sphere  are needed to correctly reproduce $\sigma_{sca}$ . This is because longitudinal isolated-sphere modes cannot be resonantly excited in homogeneous dielectric objects and, with the exception of the fundamental electric dipole, their coupling with a plane wave is weak \cite{Forestiere16,forestiere2017nanoparticle}. 


In the SI, we also show the scattered electric field in correspondence of the seven $\sigma_{sca}$ peaks (Fig. \ref{fig:SiliconZstate}). Unlike the longitudinally polarized Ag homo-dimer, it is now apparent that at any $\sigma_{sca}$ peak the near-field distributions only roughly resemble the modes that dominate the scattering response.

\begin{figure*}[h!]
\centering
	\includegraphics[width=0.65\textwidth]{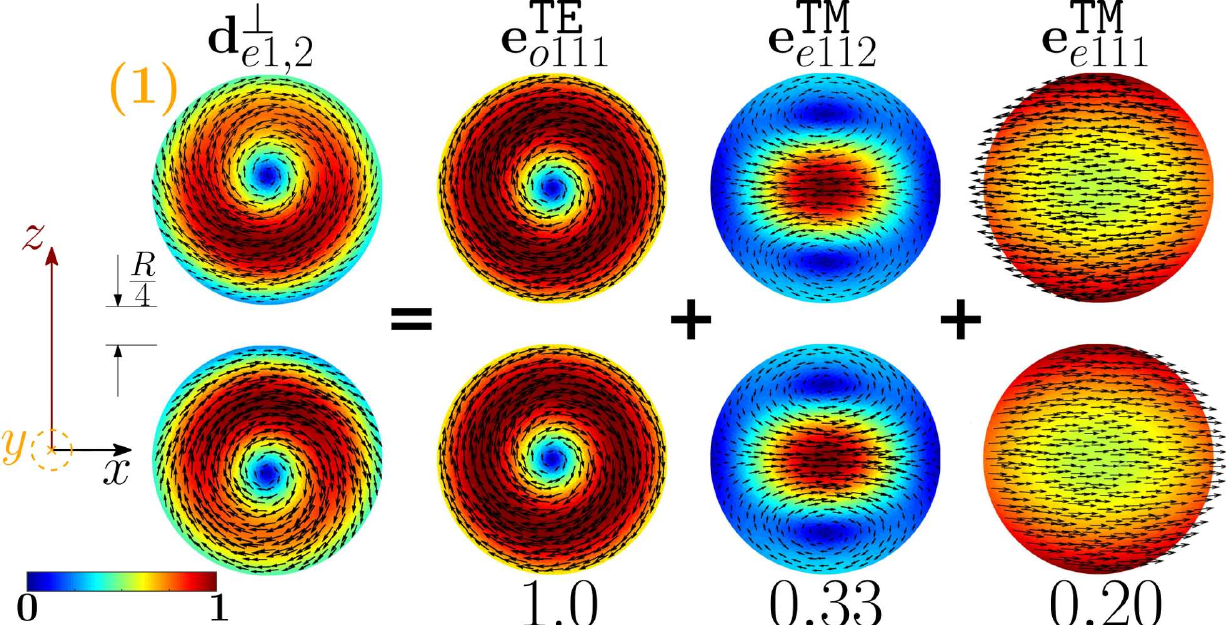}
\caption{Decomposition of the dimer-mode 
$\CCp{e1,2}$ at $x=0.771$ in terms of hybridizing isolated-sphere modes (real part of the projection on the $y=0$ plane). Each isolated-sphere modes is multiplied by the expansion coefficients of Eq. \ref{eq:HybridizationCoeff}. Below each isolated-sphere mode we also show its hybridization weight $\CoefftTM{e12\, nl}$ ($\CoefftTE{o12\, nl}$).}
\label{fig:P1SiZ}
\end{figure*}

\begin{figure}[h!]
\centering
	\includegraphics[width=0.5\textwidth]{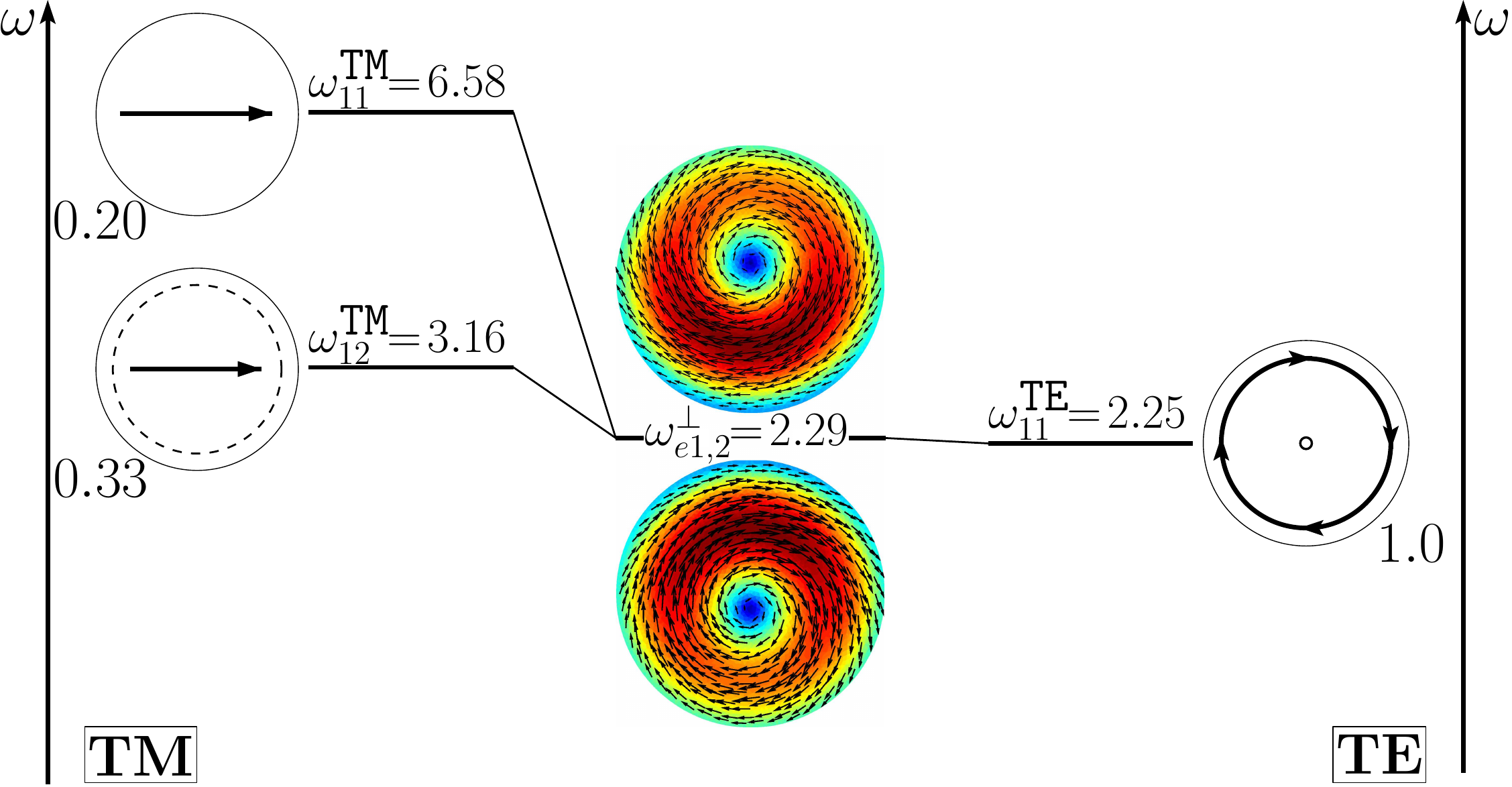}
\caption{Frequency levels describing the hybridization of the TE and TM modes of a $100$ nm isolated-sphere into the dimer-mode $\CCp{e1,2}$. The vertical axis represents the frequency (expressed in Prad/s). Next to each isolated-sphere mode we report its hybridization weight $\CoefftTM{e12\, nl}$ ($\CoefftTE{o12\, nl}$).}
\label{fig:HybDia_Siz_1}
\end{figure}
As shown in Fig. \ref{fig:P1SiZ}, the dimer-mode $\CCp{e1,2}$, which is responsible for the first peak of $\sigma_{sca}$, arises from the hybridization of the fundamental magnetic dipole $\Cb{o111}$ and the first and second order electric dipoles, i.e. $\Ca{e111}$, $\Ca{e112}$. The modes $\Ca{e111}$ and $\Ca{e112}$ constructively interfere with $\Cb{o111}$ within the region of each sphere located in between the gap and the center. The net effect is to move the vortex core away from the gap.
We also introduce in Fig. \ref{fig:HybDia_Siz_1} the {\it frequency hybridization diagram} for the dimer-mode $\CCp{e1,2}$, where  we show in the middle the dimer-mode, on the left and on the right the TE and TM  isolated-sphere modes that take part in the hybridization.  The vertical position at which both the isolated-sphere and dimer- modes are centred in the diagram is proportional to their resonant frequency obtained assuming $R=100nm$ (Tabs. \ref{tab:SiliconZ} and \ref{tab:ResonancesSiSingle} in SI). We also show their hybridization weights. 

Going back to the analysis of Fig. \ref{fig:Csca_Si_Z}, the shoulder on the right of the first peak is due to the longitudinal mode $\CCo{e0,1}$. Then we found a dip of $\sigma_{sca}$, which is due to the destructive interference between the dimer-modes $\CCp{e0,3}$ and $\CCo{e0,1}$.

The next peak, labelled with $(3)$, is mainly due to the mode $\CCp{e2,1}$, with minor contributions from the modes $\CCp{e0,3}$ and $\CCo{e0,1}$. The dimer-modes $\CCp{e2,1}$ and $\CCp{e0,3}$ arise from the dominant contributions of the fundamental magnetic quadrupole and of the second order electric dipole, respectively. 
\begin{figure*}[h!]	
\centering
	\includegraphics[width=0.85\textwidth]{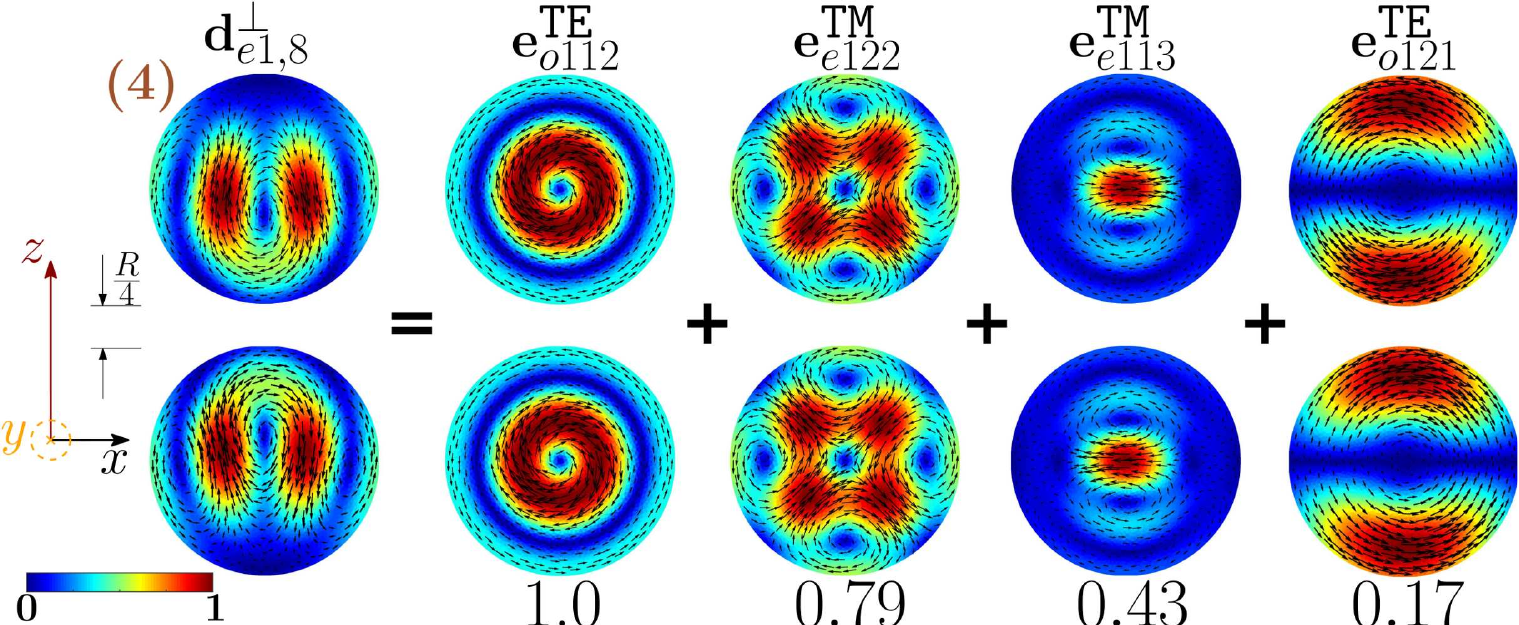}
\caption{
Decomposition of the dimer-mode 
$\CCp{e1,8}$ at $x=1.366$ in terms of hybridizing isolated-sphere modes (real part of the projection on the $y=0$ plane). Each isolated-sphere modes is multiplied by the expansion coefficients of Eq. \ref{eq:HybridizationCoeff}. Below each isolated-sphere mode we also show its hybridization weight .$\CoefftTM{e18\, nl}$ ($\CoefftTE{o18\, nl}$)}
\label{fig:P3SiZ}
\end{figure*}
\begin{figure}[h!]
\centering
	\includegraphics[width=0.5\textwidth]{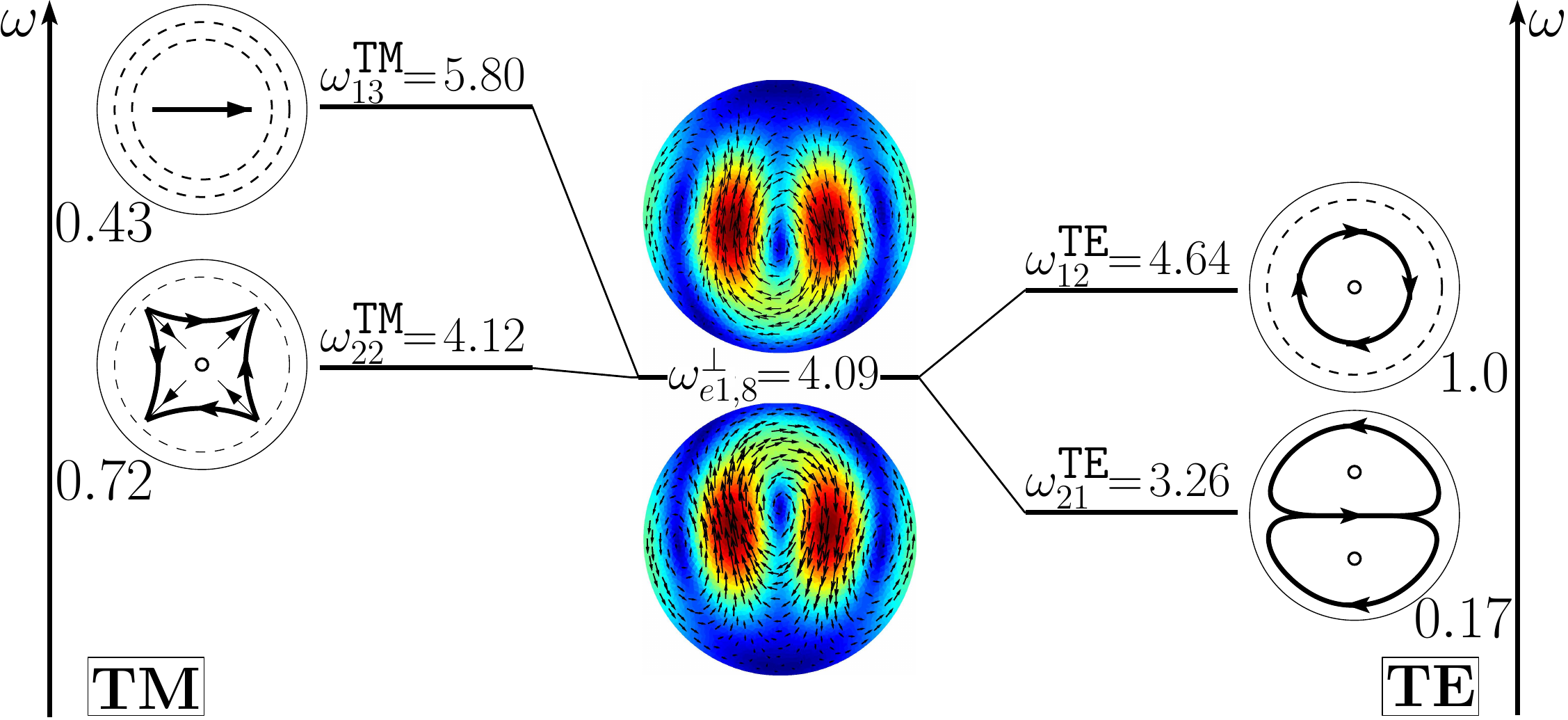}
\caption{Frequency levels describing the hybridization of the TE and TM modes of a $100$ nm isolated-sphere into the dimer-mode $\CCp{e1,8}$. The vertical axis represents the frequency (expressed in Prad/s). Next to each isolated-sphere mode we report its hybridization weight $\CoefftTM{e18\, nl}$ ($\CoefftTE{o18\, nl}$).}
\label{fig:HybDia_Siz_3}
\end{figure}

The dimer-mode behind the $\sigma_{sca}$ peak labelled with $\left( 4 \right)$ is $\CCp{e1,8}$. As shown in Fig. \ref{fig:P3SiZ}, it arises from the hybridization among the second order magnetic dipole $\Cb{o112}$, the  second order electric quadrupole $\Ca{e122}$, the third order electric dipole $\Ca{e113}$, and the fundamental magnetic quadrupole $\Cb{o121}$. In particular, $\Cb{o112}$ and $\Ca{e122}$ constructively interfere along the horizontal diameter of both spheres and destructively along the vertical diameter of both spheres.
The ``horseshoe'' shape of this dimer-mode is determined by the action of the fundamental magnetic quadrupole $\Cb{o121}$ that constructively interferes with the second order magnetic quadrupole $\Cb{o122}$ in the hemispheres of the two spheres closer to the gap, and destructively interferes in the remaining hemispheres.
The contribution of the third order electric dipole $\Ca{e113}$ pushes the vortex core within each sphere toward the gap, because it destructively interferes with the $\Cb{o112}$ in the semicircle closer to the gap, and constructively interferes in the opposite half.
In Fig. \ref{fig:HybDia_Siz_3} we also show the frequency hybridization diagram for the the dimer-mode $\CCp{e1,8}$.
\begin{figure*}[h!]
\centering
	\includegraphics[width=0.7\textwidth]{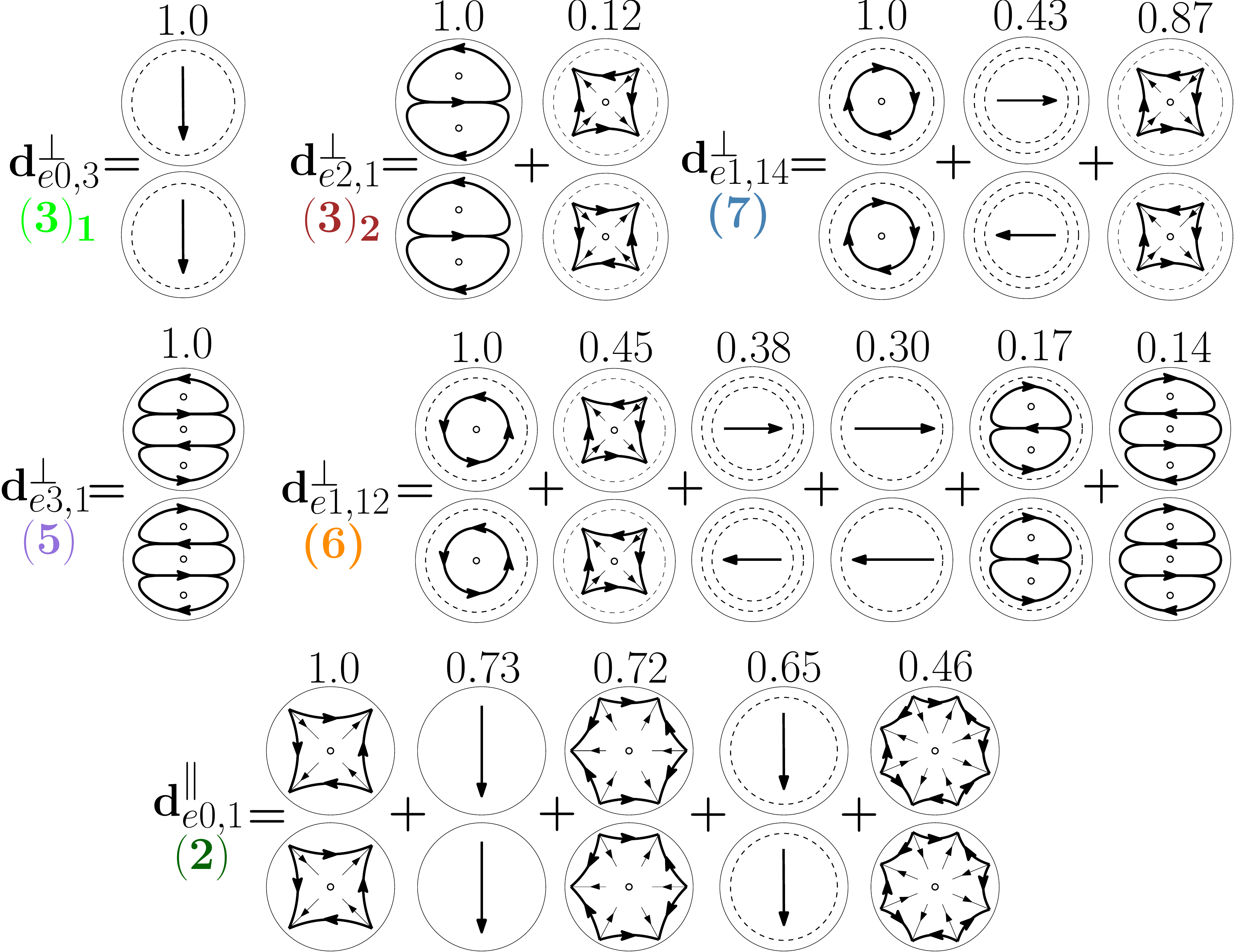}
	\caption{Decomposition of the dimer-modes  $\CCo{e0,1}$, $\CCp{e0,3}$, $\CCp{e2,1}$, $\CCp{e3,1}$, $\CCp{e1,12}$, $\CCp{e0,14}$, at $x=0.901,1.099,1.099,1.423,1.584,1.688$, respectively, in terms of hybridizing isolated-sphere modes (real part of the projection on the $y=0$ plane). Each isolated-sphere modes is multiplied by the expansion coefficients of Eq. \ref{eq:HybridizationCoeff}. Above each isolated-sphere mode we also show its hybridization weight.}
	\label{eq:SiZ_Remaining}
\end{figure*}

In Fig. \ref{eq:SiZ_Remaining} we show the hybridization diagrams of the remaining modes of Fig. \ref{fig:Csca_Si_Z}.  
Specifically, the dimer-mode $\CCo{e0,1}$ arises from the hybridization among the fundamental electric dipole $\Ca{e011}$, quadrupole $\Ca{e021}$, octupole $\Ca{e031}$, hexadecapole $\Ca{e041}$, and the second order electric dipole $\Ca{e022}$. The dimer-mode $\CCp{e0,3}$ is dominated by the second order electric dipole $\Ca{e012}$, the dimer-mode $\CCp{e2,1}$ is dominated by the fundamental magnetic quadrupole $\Cb{o221}$, the dimer-mode $\CCp{e3,1}$ is dominated by the fundamental magnetic octupole $\Cb{o331}$. There is almost no hybridization in these three cases. The mode $\CCp{e1,12}$ arises from the hybridization of the second order magnetic dipole $\Cb{o112}$,  second order electric quadrupole $\Ca{e122}$, second $\Ca{e112}$ and third order $\Ca{e113}$ electric dipoles, second order magnetic quadrupole $\Ca{e122}$, and fundamental magnetic octupole $\Cb{o131}$. Finally, the mode $\CCp{e1,14}$ results from the hybridization of second order magnetic dipole $\Cb{o122}$, third order electric dipole $\Ca{e113}$, and second order electric quadrupole $\Ca{e113}$.

As for the Ag dimer, when we increase the edge-edge distance the dimer-modes change. Nevertheless, the isolated-sphere modes used as basis set remain the same, while the hybridization weights vary.
For instance, when the edge-edge distance increases from $R/4$ to $R$ the contribution of the fundamental magnetic dipole to the dimer-mode that causes the first $\sigma_{sca}$ peak increases compared to the remaining modes (Fig. \ref{fig:SiliconZPeak1_R} in SI). This mode is the only to survive when the distance goes to infinity (Fig. \ref{fig:CscaSiSingle} in SI).
Similarly, the dimer-mode $\CCp{e1,7}$ associated to the peak $\left( 4 \right)$ (Fig. \ref{fig:SiliconZPeak4_R} in SI), arises mainly from the second order electric quadrupole $\Ca{e122}$, while the second order magnetic dipole  $\Cb{o112}$ that was dominant for edge-edge distance of $R/4$ (Fig. \ref{fig:P3SiZ}) now plays a minor role. The mode $\Ca{e122}$ is the only to survive when the distance goes to infinity (Fig. \ref{fig:CscaSiSingle}).

%% file: SiX_R4/SiX.tex
\subsection*{Transversely Polarized Si homo-dimer}
We now still study the modes excited by a silicon homo-dimer, but excited by a plane-wave propagating along the dimer's axis direction $\hat{\bf z}$ and polarized along the transverse $\hat{\bf x}$-direction.
\begin{figure*}[h!]
\centering
\includegraphics[width=1\textwidth]	{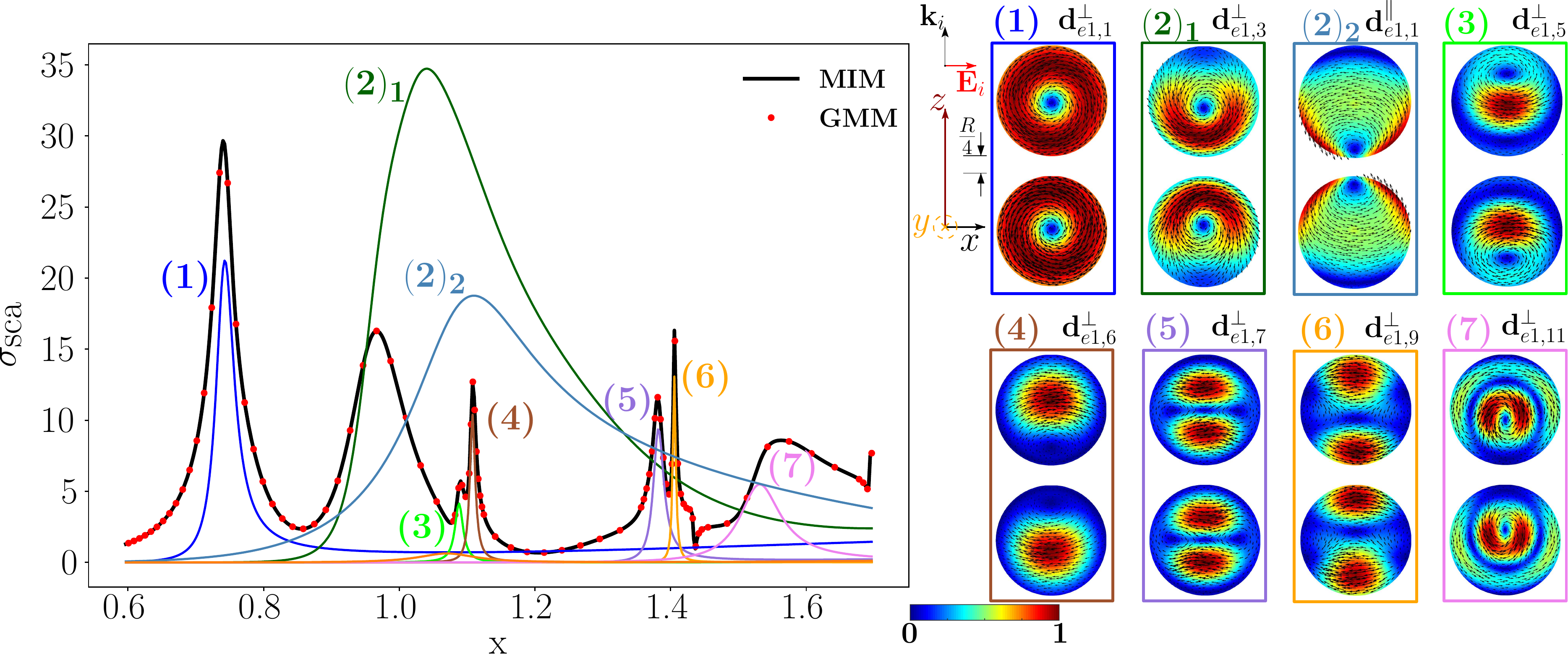}
\caption{Scattering efficiency $\sigma_{sca}$ of a Si-spheres homo-dimer as a function of the spheres size parameter $x=2\pi R/\lambda$, obtained via the material-independent-mode expansion (black line) and by direct-calculation (red dots). The radius of each sphere is $R$, the edge-edge distance is $R/4$. The dimer is excited by a plane wave propagating along the dimer's axis $\hat{\bf z}$ and polarized along the transverse direction $\hat{\bf x}$. Partial scattering efficiency (in color) of eight dominant dimer-modes whose real part projections on the $y=0$ plane are shown on the right.}
\label{fig:Csca_Si_X}
\end{figure*}
In Fig. \ref{fig:Csca_Si_X}, we plot the scattering efficiency obtained by the material-independent-mode expansion of Eq. \ref{eq:DimerModeExpansion} (black line), and by the direct GMM calculation \cite{xu95} (red dots) as a function of the size parameter $x$.
\begin{table*}[h!]
\centering
\begin{tabular}{|c||cccc||cc|}
\hline
 index & $x^\nu_{e1q}$ & $\omega_{e1q}^\nu$ & $\varepsilon^\nu_{e1q}$ & $\rho^\nu_{e1q}$ & \# peak & $x$ \\
\hline
$\CCp{e1,1} $ & 0.742 & 2.22 & 16.01-0.74i & 0.049 &(1) & 0.740 \\
$\CCp{e1,3} $ & 0.960 & 2.88 & 16.96-2.70i & 0.191 &$(2)$ & 0.967 \\
$\CCo{e1,1}$ & 1.326 & 3.98 & 1.28 - 2.64i & 0.997& $(2)$ & 0.967 \\
$\CCp{e1,5} $ & 1.088 & 3.26 & 15.99-0.22i & 0.015 &(3) & 1.090 \\
$\CCp{e1,6} $ & 1.108 & 3.32 & 15.99-0.14i & 0.009 &(4) & 1.108 \\
$\CCp{e1,7} $ & 1.381 & 4.14 & 16.02-0.24i & 0.016 &(5) &  1.381 \\
$\CCp{e1,9}$ & 1.406  & 4.21 & 15.99-0.06i & 0.004 &(6) & 1.406 \\
$\CCp{e1,11}$ & 1.531 & 4.58 & 15.99-0.79i & 0.052 &(7) & 1.560 \\
\hline
\end{tabular}
\caption{Resonant size parameter $x$, corresponding values of resonant frequency,  eigenpermittivity, and residuum of the dimer-modes which dominate the scattering efficiency of Fig. \ref{fig:Csca_Si_X}. Positions of the peaks of the total scattering efficiency.}
\label{tab:SiX}
\end{table*}
We also show in color the partial scattering efficiency of the eight dominant dimer-modes, whose real projections on the $y=0$ plane are represented on the right. 
It is worth to point out that by calculating $\sigma_{sca}$ using only these $8$ dimer-modes, the agreement with the GMM remains satisfactory (Fig. \ref{fig:Csca_Si_X_OPh} of SI).
We list in Tab. \ref{tab:SiX} the resonant size parameter $x$ of the eight dominant modes, the corresponding resonant frequency assuming $R=100$ nm, and the corresponding residuum. 
{
The residuum of the longitudinal mode $\CCo{e1,1}$ is almost one-order of magnitude higher than the residuum of the transverse modes in Tab. \ref{tab:SiX}.  This fact can be again  interpreted by Eq. \ref{eq:ResonantCondition}: the real part of the eigen-permittivity $\varepsilon_{e1,1}^\parallel \left( x \right)$ never equates the Si eigen-permittivity, since irrespectively of $x$, we have 	 $ -5.37 <\varepsilon_{e1,1}^\parallel \left( x \right) <1.36$ and $\varepsilon_{R,Si}=16$.  On the contrary, for the transverse modes there always exists  a value of $x$ in correspondence of which $\re{\varepsilon_{emq}^\perp \left( x \right)} = \varepsilon_{R,Si}$, as it also apparent from the fourth column of Tab. \ref{tab:SiX}. Nevertheless, the mode $\CCo{e0,1}$ has to be considered because it strongly couples with the plane wave and it also has  a stronger radiative strength \cite{forestiere2017nanoparticle}.
A low residuum corresponds to a sharp scattering peak: the mode $\CCp{e1,9}$ exhibits the lowest residuum and it also associated to the sharpest $\sigma_{sca}$ peak.  

In the SI, we also show the near field pattern of the scattered electric fields on the $y=0$  plane of the spheres in correspondence of the eight peaks (Fig. \ref{fig:SiliconXstate}).


\begin{figure}[h!]
\centering
	\includegraphics[width=0.5\textwidth]
	{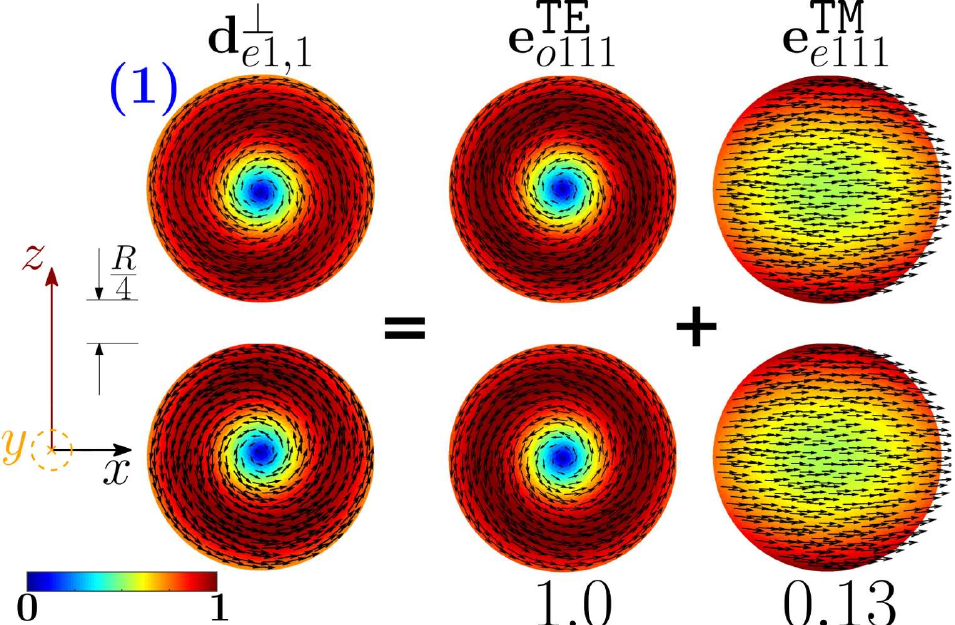}
\caption{Decomposition of the dimer-mode $\CCp{e1,1}$ at $x=0.740$, in terms of hybridizing isolated-sphere modes (real part of the projection on the $y=0$ plane). Each isolated-sphere modes is multiplied by the expansion coefficients of Eq. \ref{eq:HybridizationCoeff}. Below each isolated-sphere mode we also show its hybridization weight $\CoefftTM{e11\, nl}$ ($\CoefftTE{o11\, nl}$).}
\label{fig:SiX_P1}
\end{figure}

\begin{figure}[h!]
\centering
	\includegraphics[width=0.5\textwidth]{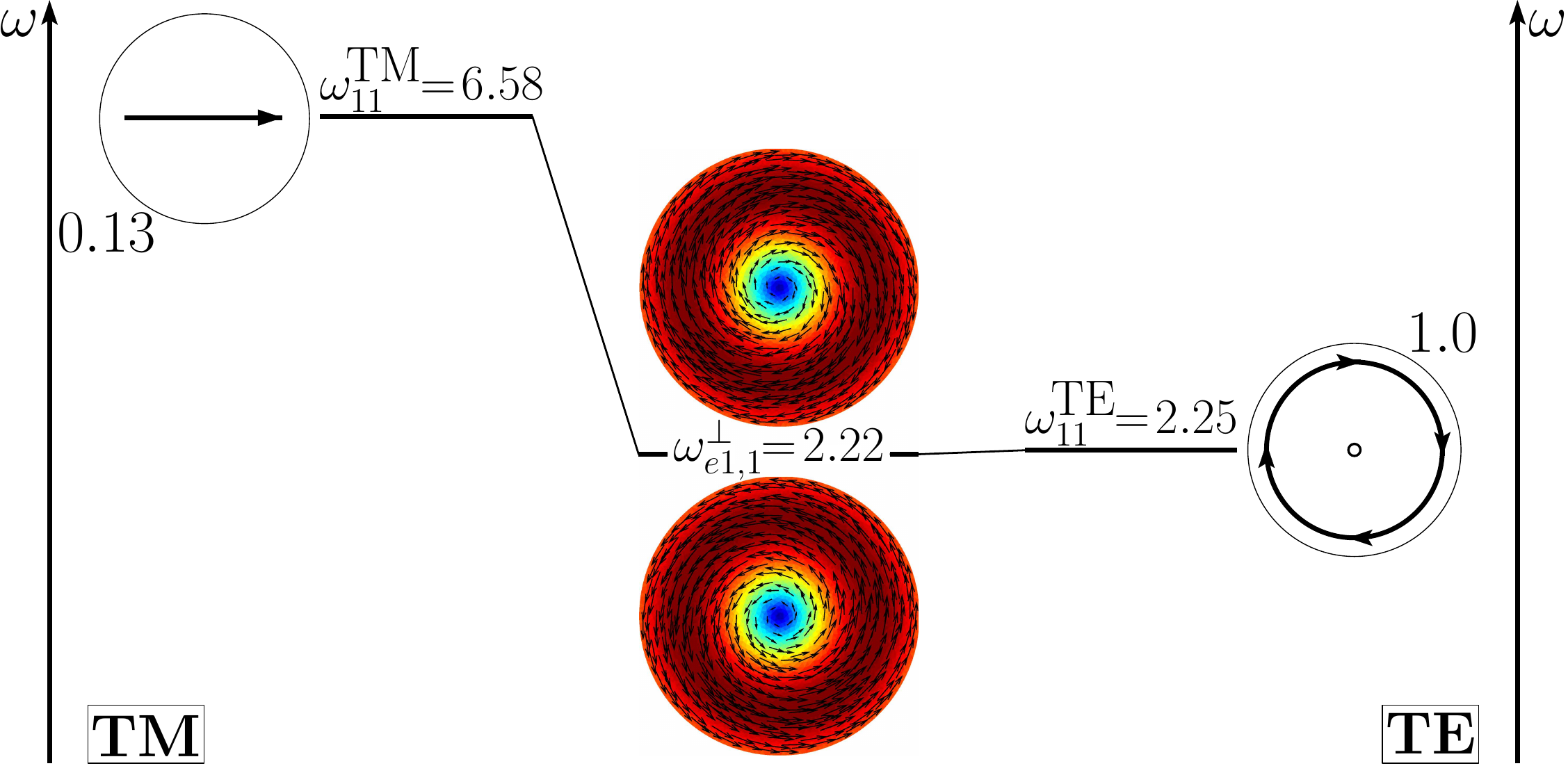}
\caption{Frequency levels describing the hybridization of the TE and TM modes of a $100$ nm isolated-sphere into the dimer-mode $\CCp{e1,1}$. The vertical axis represents the frequency (expressed in Prad/s). Next to each isolated-sphere mode we show its hybridization weight $\CoefftTM{e11\, nl}$ ($\CoefftTE{o11\, nl}$).}
\label{fig:Si_X_HybDia_1}
\end{figure}

The mode $\CCp{e1,1}$ causes the first peak of the $\sigma_{sca}$ spectrum; in Fig. \ref{fig:SiX_P1} it is decomposed in terms of hybridizing isolated-sphere modes. The mode $\CCp{e1,1}$ almost coincides with the isolated-sphere fundamental magnetic dipole apart from a small contribution from the fundamental electric dipole. In  in Fig. \ref{fig:Si_X_HybDia_1} we also show the corresponding ``frequency'' hybridization diagram. 

\begin{figure}[h!]
\centering
	\includegraphics[width=\textwidth]{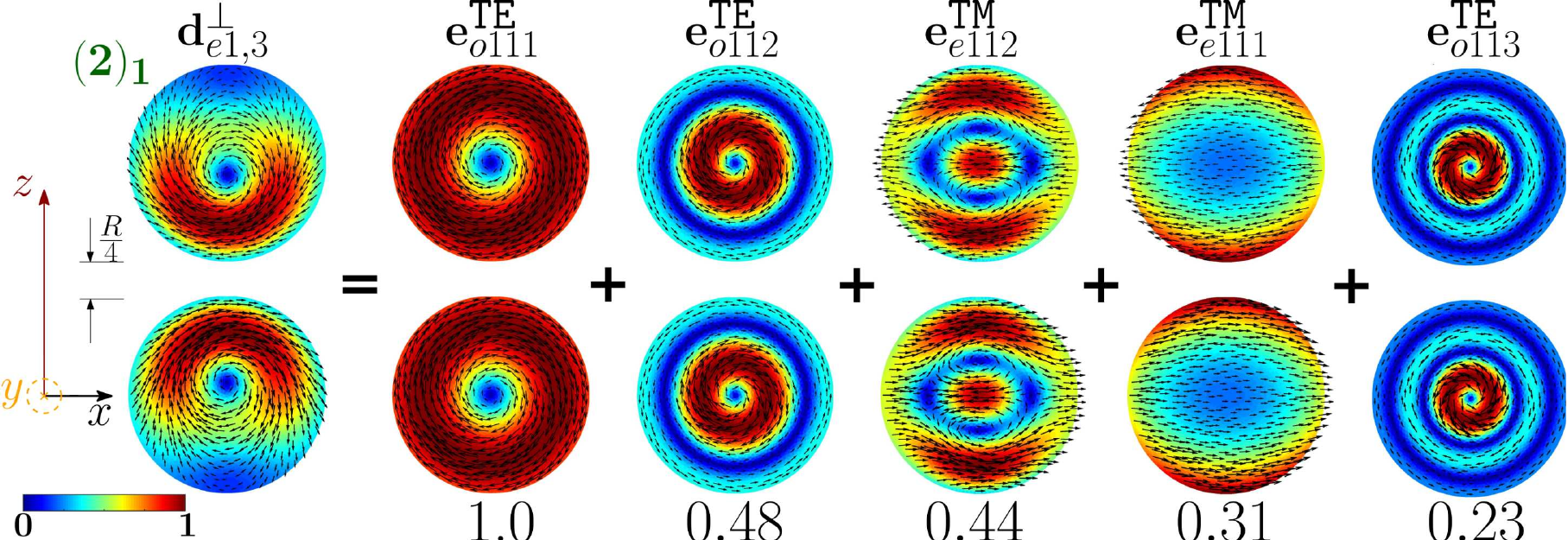}
\caption{Decomposition of the dimer-mode $\CCp{e1,3}$ at $x=0.967$, in terms of isolated-sphere modes  (real part of the projection on the $y=0$ plane). Each isolated-sphere modes is multiplied by the expansion coefficients of Eq. \ref{eq:HybridizationCoeff}. Below each isolated-sphere mode we show its hybridization weight $\CoefftTM{e13\, nl}$ ($\CoefftTE{o13\, nl}$).}
\label{fig:SiX_P2}
\end{figure}

The second peak of the $\sigma_{sca}$ spectrum is dominated by the dimer-mode $\CCp{e1,3}$. As shown in Fig. \ref{fig:SiX_P2}, the spatial distribution of $\CCp{e1,3}$ on the $y=0$ plane originates from the interference between the fundamental magnetic dipole $\Cb{o111}$ and the two electric dipoles $\Ca{e111}$, $\Ca{e112}$ that enhance the field in the two hemispheres closer to the gap and attenuate it in the two opposite hemispheres.
In Fig. \ref{fig:Si_X_HybDia_2} we show the corresponding ``frequency'' hybridization diagram.

\begin{figure}[h!]
\centering
	\includegraphics[width=0.5\textwidth]{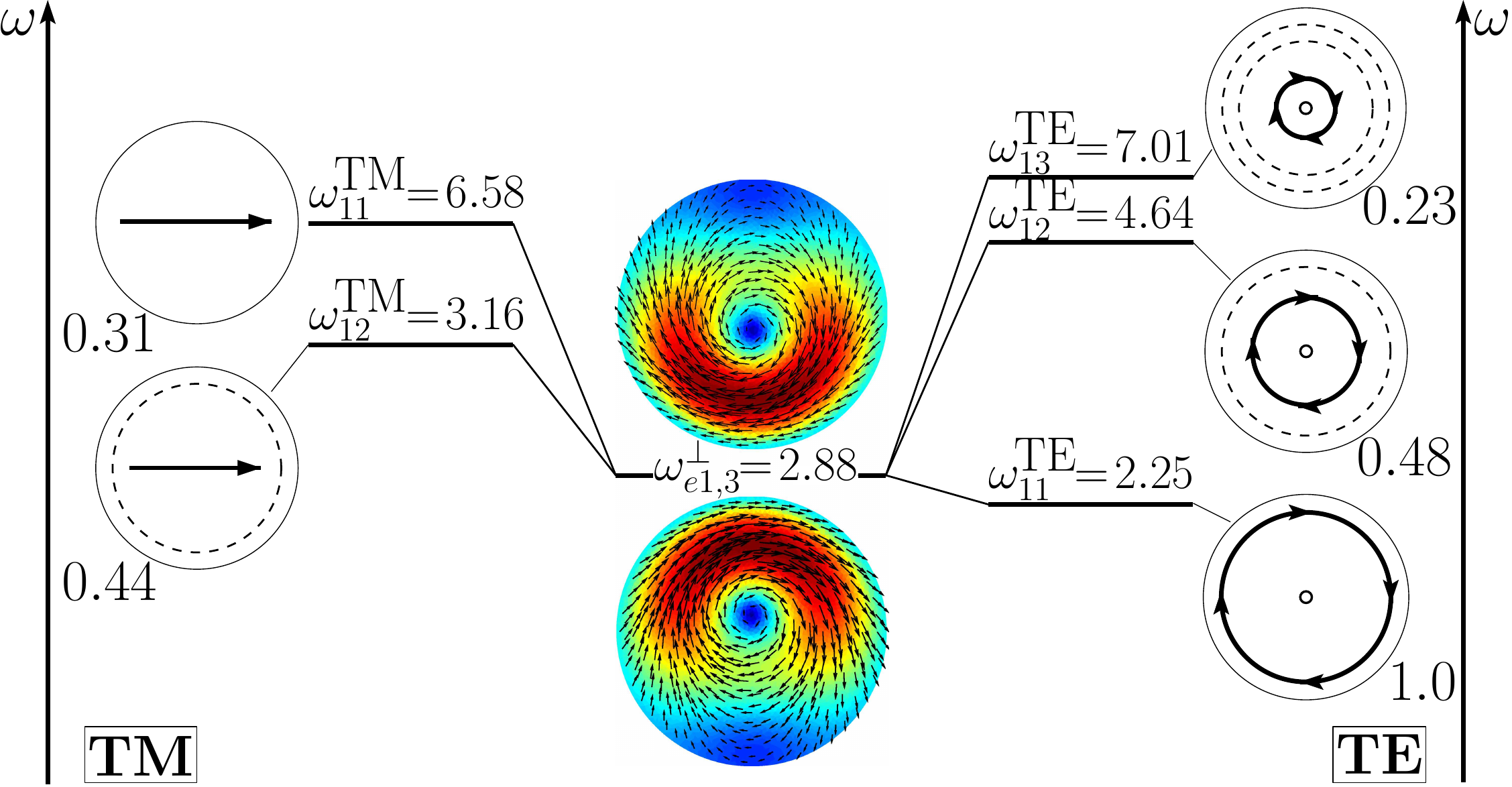}
\caption{Frequency levels describing the hybridization of the TE and TM modes of a $100$ nm isolated-sphere into the mode $\CCp{e1,3}$ of a homo-dimer with edge-edge separation of $25nm$. The vertical axis represent the frequency. Next to each isolated-sphere mode we show its hybridization weight $\CoefftTM{e13\, nl}$ ($\CoefftTE{o13\, nl}$).}
\label{fig:Si_X_HybDia_2}
\end{figure}

\begin{figure}[h!]
\centering
	\includegraphics[width=0.9\textwidth]{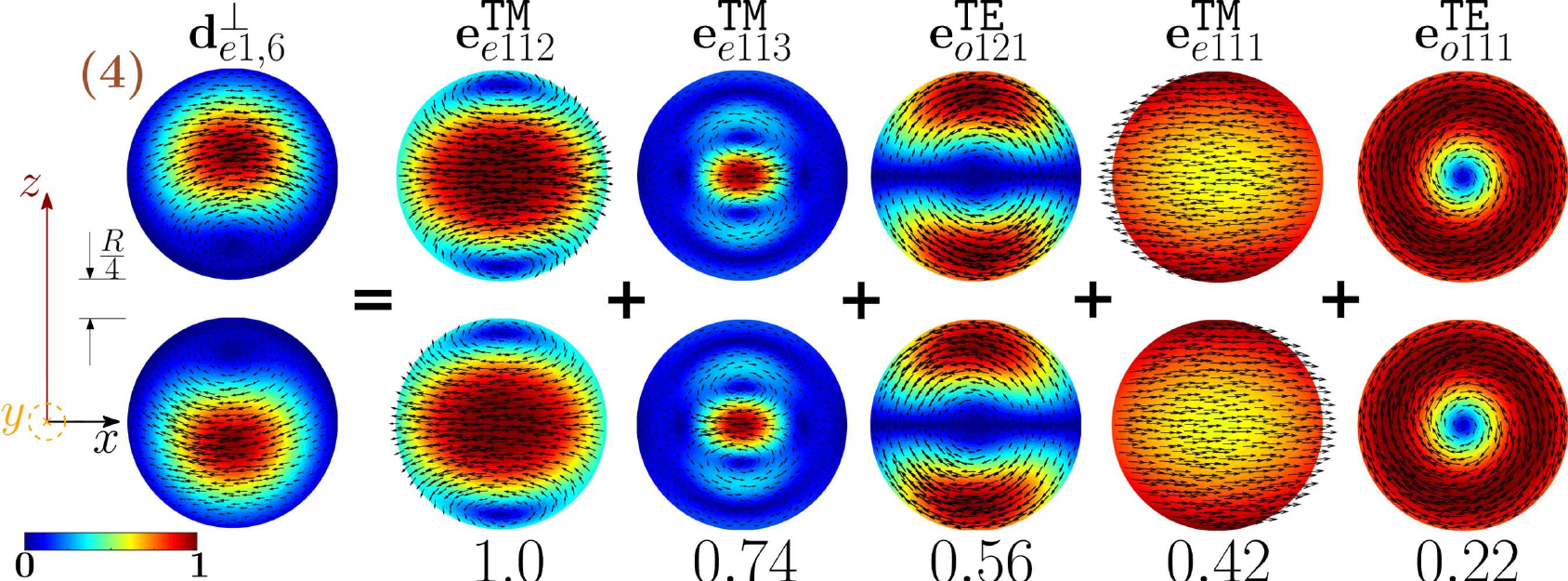}
\caption{Decomposition of the dimer-mode $\CCp{e1,6}$ at $x=1.108$,  in terms of isolated-sphere modes Each isolated-sphere modes is multiplied by the expansion coefficients of Eq. \ref{eq:HybridizationCoeff}. Below each isolated-sphere mode we show its hybridization weight $\CoefftTM{e16\, nl}$ ($\CoefftTE{o16\, nl}$).}
\label{fig:SiX_P4}
\end{figure}

The fourth peak of the $\sigma_{sca}$ spectrum is ascribed to the dimer-mode $\CCp{e1,6}$. This mode results from the hybridization of the isolated-sphere modes shown in Fig. \ref{fig:SiX_P4}. Although the second order electric dipole $\Ca{e112}$ is dominant, the contributions of $\Cb{o121}$ and $\Cb{o111}$ are very significant: they positively interfere with $\Ca{1112}$ in the two hemispheres opposite to the gap. This fact determines the shift of the mode maximum away from the gap. We also show in Fig. \ref{fig:Si_X_HybDia_4} the corresponding ``frequency'' hybridization diagram of $\CCp{e1,6}$.

\begin{figure}[h!]
\centering
	\includegraphics[width=0.5\textwidth]{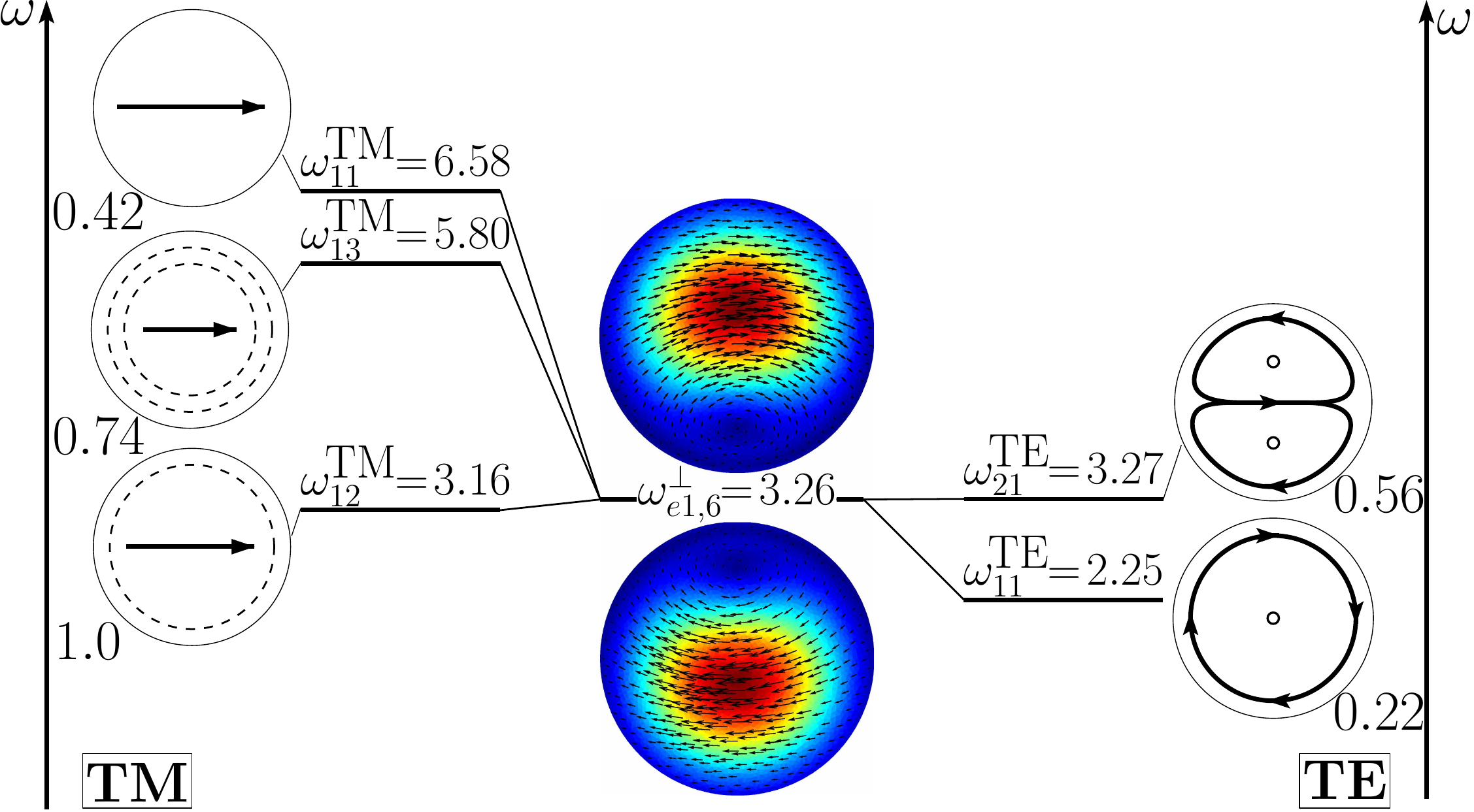}
\caption{Frequency levels describing the hybridization of the TE and TM modes of a $100$ nm isolated-sphere into the mode $\CCp{e1,6}$ of a homo-dimer with edge-edge separation of $25nm$. The vertical axis represents the frequency (expressed in Prad/s). Next to each isolated-sphere mode we show its hybridization weight $\CoefftTM{e16\, nl}$ ($\CoefftTE{o16\, nl}$).}
\label{fig:Si_X_HybDia_4}
\end{figure}

\begin{figure}[h!]
\includegraphics[width=0.9\textwidth]{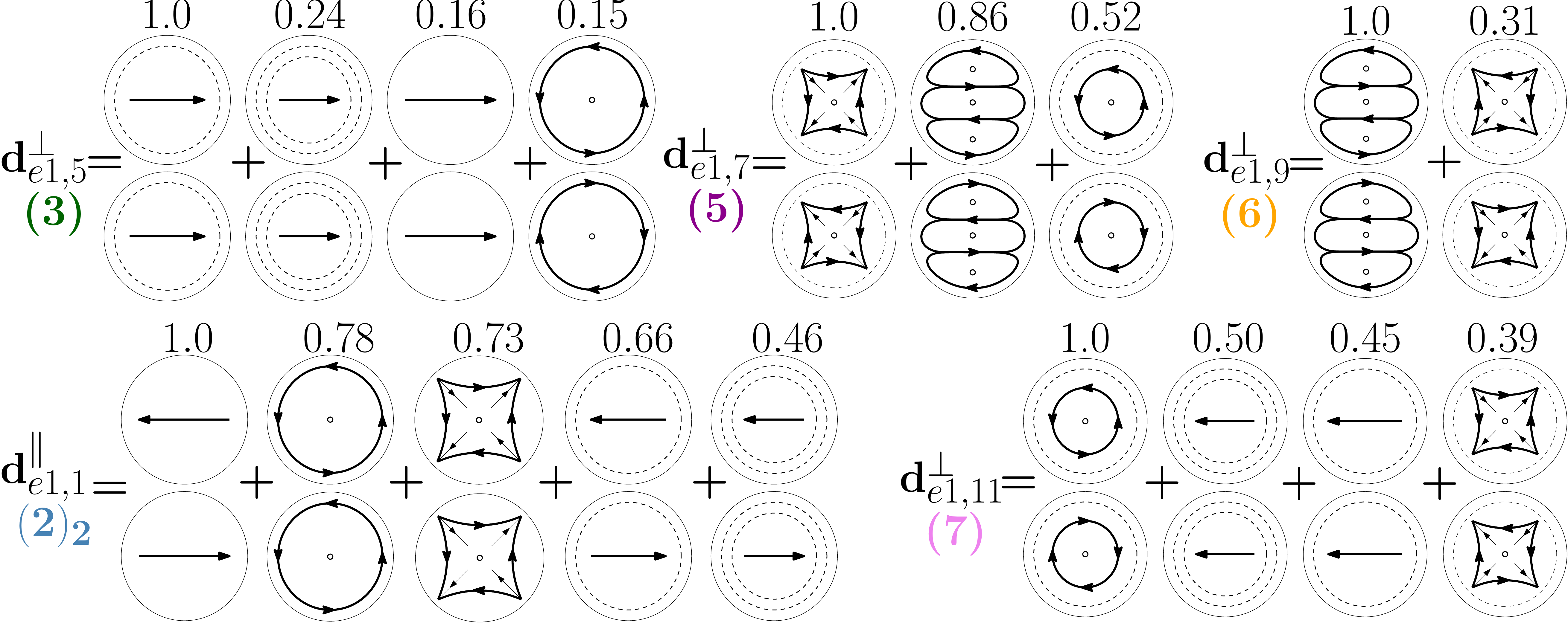}
	\caption{Decomposition of the dimer-modes  $\CCp{e1,1}$, $\CCo{e1,1}$, $\CCp{e1,7}$, $\CCp{e1,9}$, $\CCp{e1,11}$, at $x=1.099,1.099,1.423,1.584,1.688$, respectively, in terms of hybridizing isolated-sphere modes (real part of the projection on the $y=0$ plane). Each isolated-sphere modes is multiplied by the expansion coefficients of Eq. \ref{eq:HybridizationCoeff}. Above each isolated-sphere mode we also show its hybridization weight.}
	\label{fig:SiX_Remaining}
\end{figure}

In Fig. \ref{fig:SiX_Remaining} we show the decomposition in terms of isolated-sphere modes of the remaining dimer-modes dominating the far field scattering response. Specifically, the dimer-mode $\CCo{e1,1}$ is due to the hybridization of the fundamental $\Ca{e111}$, second $\Ca{e112}$ and third order $\Ca{e113}$ electric dipoles, fundamental magnetic dipole $\Cb{o111}$, and fundamental electric quadrupole $\Ca{e141}$. Furthermore,   $\CCp{e1,5}$ is due to the second order electric dipole $\Ca{1112}$ and to a lesser extent to the fundamental $\Ca{e111}$ and third order $\Ca{e113}$ electric dipole and to the fundamental magnetic dipole $\Cb{o111}$. The dimer-mode $\CCp{e1,7}$ arises from the hybridization of the second order electric quadrupole $\Ca{e122}$, the fundamental magnetic octupole $\Cb{e131}$, and the second order magnetic dipole $\Cb{o112}$.
Mode $\CCp{e1,9}$ results from the combination of the fundamental magnetic octupole $\Cb{o131}$ and the second order electric quadrupole $\Ca{e122}$. Mode $\CCp{e1,11}$ is due to the hybridization of the second order magnetic dipole $\Cb{o112}$, the second $\Ca{e112}$ and third order $\Ca{e113}$ electric dipoles, and of the second order electric quadrupole $\Ca{e122}$.

Also in this case we repeated the mode decomposition of the scattering efficiency for an edge-edge distance between the spheres equal to $R$ (Fig. \ref{fig:Csca_Si_X_R} in SI). The fundamental magnetic dipole mode ${\bf e}^\mathtt{TE}_{o111}$ becomes more and more dominant in the hybridization of the first peak (Fig. \ref{fig:SiliconXPeak1_R}).
The second peak arises from the positive interference between two dimer-modes, which are dominated by the fundamental and second order electric dipoles, respectively (Fig. \ref{fig:SiliconXPeak2_R}).
The second order electric dipole and the fundamental magnetic quadrupole become dominant in the hybridization of the third peak (see Fig. \ref{fig:SiliconXPeak3_R}). 
The fourth, fifth and sixth peaks are dominated by the isolated-sphere modes ${\bf e}^\mathtt{TM}_{e122}$, ${\bf e}^\mathtt{TE}_{o131}$, ${\bf e}^\mathtt{TE}_{o112}$, respectively.
These are the modes that survive when the edge-edge distance become infinity, as it is apparent from the $\sigma_{sca}$ of the isolated Si sphere (Fig. \ref{fig:CscaSiSingle} in the SI).

%% file: Source/Conclusions.tex
We have investigated the modes and resonances in the electromagnetic scattering from a dimer of spheres by using the full-Maxwell equations and the material independent modes. The electromagnetic scattering response of the dimer is described by a set of dimer-modes. Each dimer-mode is seen as the result of the hybridization of the modes of two constituent spheres, whose importance is quantified by hybridization weights.  As we vary the gap size, although the dimer-modes change, they are still represented in terms of the same set of isolated-sphere modes, but with different hybridization weights. This study represents the first full-Maxwell theory of hybridization in Si dimers, and it also constitutes an extension of the plasmon-mode hybridization theory to the full-retarded scenario. The  modes are classified accordingly to their behaviour in the long-wavelength limit: the longitudinal modes are the ones that become irrotational everywhere in long-wavelength limit (quasi-stationary electric modes or plasmonic modes); the transverse modes are the ones that become solenoidal everywhere in the long-wavelength limit (quasi-stationary magnetic modes or photonic modes). The transverse modes arise from the magnetic interaction, while the longitudinal modes arise from the electrical interaction.  Transverse dimer-modes cannot be resonantly excited in metal dimers. In general, the longitudinal and the transverse modes of the dimer arise from the hybridization of both the longitudinal and the transverse modes of the isolated-sphere. The retardation effects play an important role in the mode-coupling with the incident field. Radiation losses may lead to a significant broadening of the resonance peaks. 

By using this theoretical framework, we investigate the resonant scattering from metal and dielectric dimers with dimensions of the order of the incident wavelength under different plane-wave illuminations  and different gap sizes. The scattering efficiencies can be accurately described by a very limited number of dimer-modes. Then, we quantitatively decompose these dimer-modes into the modes of the constituent spheres, providing the corresponding hybridization weights. We also study how the hybridization weights change as the gap size is varied.
The longitudinal dimer-modes are sufficient to describe the far-field scattering from metal dimers. On the contrary, the far field scattering from dielectric dimers involve resonant transverse dimer-modes and off-resonance longitudinal dimer-modes.

 To offer an intuitive understanding we extend the hybridization diagrams introduced by Prodan et al. to the full-Maxwell analysis, showing the hybridization of the electric and magnetic modes of an isolated Si sphere into the dimer-modes and the corresponding frequency levels.

%% file: Source/Methods.tex
In this {\it methods} section, we show the derivation of the modes of a sphere  dimer, by representing each dimer-mode in terms of a weighted combination, i.e. hybridization, of isolated-sphere modes. We also provide the explicit expressions for the corresponding hybridization coefficients given in Eq. \ref{eq:HybridizationCoeff}. 

At the very basis of any extension of the Mie theory to sphere dimers lies the VSWF addition theorem. This theorem enables the representation of the radiating VSWFs centred on one origin as an expansion of regular VSWFs centred on a different origin. The addition theorem was first derived in Refs.  \cite{stein1961addition,cruzan1962translational}, and  it was later combined with the Mie theory in Refs. \cite{liang1967scattering,bruning1971multiple}. Subsequently, it was significantly improved by many authors including Borghese et al. \cite{borghese1979electromagnetic}, Fuller et al. \cite{fuller1988consummate}, and Mackowski \cite{mackowski1991analysis,mackowski1994calculation}. Very detailed introductions can be found in Refs. \cite{mishchenko1999light,borghese2007scattering}

Let us consider the problem of scattering by a dimer of spheres in free-space. The geometry of the problem is sketched in Fig. \ref{fig:Domain_dimer}. The spheres have radius $R_1$ and $R_2$, respectively, and they occupy the regions $\Omega_1$ and $\Omega_2$, while the surrounding space is denoted with $\Omega_3$. The sphere $\Omega_1$ is centred on the origin of a Cartesian coordinate system $O_1 {\bf r}_1$ , while the sphere $\Omega_2$ is centred on the origin of a second Cartesian coordinate system $O_2 {\bf r}_2$. The coordinate system $O_2 {\bf r_2}$ is obtained by translating the coordinate system $O_1 {\bf r_1}$ through a distance $D$ along the $z$ axis. The dimer is aligned along the direction of the $z$-axis of both coordinate systems. The two spheres have the same material composition, which is assumed to be linear, non-magnetic, isotropic, homogeneous in time and space, non dispersive in space, and time-dispersive with relative permittivity $\varepsilon_{R} \left( \omega \right)$.  
The object is excited by a time harmonic electromagnetic field incoming from infinity $\re{\Ei  e^{- i \omega t}}$.
We denote the total electric field as $ { \bf E }$, while the scattered electric field is denoted as ${\bf E}_S = {\bf E} - {\bf E}_i$. The scattered field ${\bf E}_S$ is the solution of the following problem:  

\begin{equation}
\begin{aligned}
\label{eq:MErot12} 
& k_0^{-2} \boldsymbol{\nabla}^2 \Es +  \varepsilon_{R}\left( \omega \right)   \Es = \left[1- \varepsilon_{R} \left( \omega \right)  \right]\Ei \qquad&& \mbox{in} \, \Omega_j, \,\,j\in \{1,2\} \\
& k_0^{-2} \boldsymbol{\nabla}^2 \Es + \Es = {\bf 0}  \; && \mbox{in} \, \Omega_3, 
 \end{aligned}
\end{equation}

\begin{equation}
\begin{aligned}
\label{eq:BC1}
   & \rvers_j \times \left( \Es|_{\Omega_3} - \Es|_{\Omega_j} \right) = {\bf 0} \\
   & \rvers_j \times \left( \boldsymbol{\nabla} \times \Es |_{\Omega_3} - \boldsymbol{\nabla} \times \Es |_{\Omega_j} \right) = {\bf 0}
   \end{aligned} \qquad
    \, \mbox{on}\; r_j=R_j,\,\, j\in\{1,2\} \\
\end{equation}
where $k_0= \omega/c_0$, $\rvers_j$ is the radial versor of the reference frame centered in the $j$-th sphere. Equations \ref{eq:MErot12}-\ref{eq:BC1} have to be solved with the  radiation conditions at infinity. 

Aiming at the reduction of the scattering problem to
an algebraic form, we introduce the auxiliary homogeneous problem that is obtained from Eqs. \ref{eq:MErot12}-\ref{eq:BC1} by zeroing the driving term ${\bf E}_i$:
\begin{equation}
\label{eq:AuxHelmholtz}
\begin{aligned}
& -{k_0^{-2}} \, \boldsymbol{\nabla}^2 {\bf C}  =  \gamma {\bf C} \qquad  && \mbox{in} \; \Omega_1 \cup \Omega_2, \\
& -k_0^{-2} \boldsymbol{\nabla}^2 {\bf C} = {\bf C}   \; \qquad && \mbox{in} \, \Omega_3, 
 \end{aligned}
\end{equation}
\begin{equation}
\begin{aligned}
   & \rvers_j \times \left( {\bf C}|_{\Omega_3} - {\bf C}|_{\Omega_j} \right) = {\bf 0} \\
   & \rvers_j \times \left( \boldsymbol{\nabla} \times {\bf C} |_{\Omega_3} - \boldsymbol{\nabla} \times {\bf C} |_{\Omega_j} \right) = {\bf 0}
   \end{aligned} \qquad
    \, r_j=R_j,\,\, j\in\{1,2\}, \\
    \label{eq:BC1_h}
\end{equation}
with the radiation condition at infinity, 
where $\gamma$ is the eigenvalue and ${\bf C}\left( {\bf r}\right)$ is the corresponding eigenfunction. 

Exploiting the symmetry of the problem, we can expand the vector field ${\bf C}$, solution of the homogeneous problem of Eqs. \ref{eq:AuxHelmholtz} and \ref{eq:BC1_h} in terms of two basis set, namely the material-independent-modes of the two isolated spheres \cite{Forestiere16} as:
\begin{equation}
 {\bf C} =  \sum_{j=1}^2 \sum_{p\in\left\{e,o\right\}} \sum_{m=0}^{\infty} {\bf C}^{\left( j \right)}_{pm},
 \label{eq:Cdimer}
\end{equation}
where ${\bf C}^{\left( j \right)}$ in the reference system $O_j {\bf r}_j$ assumes the following form:
\begin{equation}
{\bf C}^{\left( j \right)}_{pm} \left( {\bf r}_j \right) = 
\displaystyle\sum_{ n l}
\left[
\uTM{pmnl}{j} \Cah{pmnl}({\bf r}_j)
+
\uTE{\bar{p}mnl}{j} \Cbh{\bar{p}mnl}({\bf r}_j)
\right],
\label{eq:Single-sphere-scatt1}
\end{equation}
$\displaystyle\sum_{nl}=\displaystyle\sum_{n=\max\left(1,m \right)}^{\infty}\displaystyle\sum_{l=1}^\infty$,
 $(\bar{\cdot})$ is the binary operator defined as $ \bar{e} = o $ and $\bar{o} = e $, and $\left\{\Caj{pmnl},\,\Cbj{pmnl}\right\}$ are the isolated-sphere modes of the $j$-th sphere	. They have the form:
\begin{equation}
\begin{aligned}
\Cah{pmnl}({\bf r}_j)& = 
 \left\{ 
  \begin{array}{cl}
 \Ni{1}{ \SqrtTM k_0 {\bf r}_j }{pmn} & 
 {\bf r}_j \in \Omega_j, \\
 \inoutTM{nl}{j} \Ni{3}{ k_0 {\bf r}_j}{pmn} & 
 {\bf r}_j \in \mathbb{R}^3 \backslash {\Omega}_j,
  \end{array}
   \right.\\
\Cbh{pmnl}({\bf r}_j) & = \left\{ 
  \begin{array}{cl}
   \Mi{1}{\SqrtTE k_0 {\bf r}_j}{ pmn} & {\bf r}_j \in \Omega_j, \\
  \inoutTE{nl}{j} \Mi{3}{ k_0 {\bf r}_j}{pmn} & {\bf r}_j \in \mathbb{R}^3 \backslash \Omega_j.
  \end{array} \right.,
\end{aligned}
\label{eq:IsolatedEigVec}
\end{equation}
where
\begin{equation}
\begin{aligned}
\inoutTM{nl}{j} &= \SqrtTM  \frac{\J{\SqrtTM  x_j}{n}}{h_n^{\left(1\right)}\left( x_j\right)}, \\ \inoutTE{nl}{j} &= \frac{\J{\SqrtTE  x_j}{n}}{h_n^{\left(1\right)}\left( x_j\right)}. 
\end{aligned}
\end{equation}
The functions $\left\{ {\bf M}^{(1)}_{pmn},{\bf N}^{(1)}_{pmn}\right\}$ and  $\left\{ {\bf M}^{(3)}_{pmn},{\bf N}^{(3)}_{pmn}\right\}$ are the regular and radiating vector spherical wave functions (VSWFs), whose radial dependence is given by the spherical Bessel $j_n$ and Hankel $h_n^{\left( 1 \right)}$ functions of the first kind, respectively. The subscript $p \in \left\{ e, o \right\}$ denote even and odd azimuthal dependence. The quantities $\eigTMl_{nl}$ and $\eigTEl_{nl}$ are eigen-permittivities of the $j$-th isolated-sphere.
By using Eqs. \ref{eq:IsolatedEigVec} into \ref{eq:Single-sphere-scatt1} we obtain:
\begin{equation}
{\bf C}_{pm}^{\left( j \right)} \left( {\bf r}_j \right) = \left\{
  \begin{array}{ll}
\displaystyle\sum_{ n l}
\left[
\uTM{pmnl}{j} \Ni{1}{\SqrtTM k_0 {\bf r}_j}{pmn}
+
\uTE{\bar pmnl}{j} \Mi{1}{\SqrtTE k_0 {\bf r}_j}{\bar pmn}
\right] \quad &{\bf r}_j\in \Omega_j, \\
\displaystyle\sum_{ n l}
\left[
\uTM{pmnl}{j} \inoutTM{nl}{j} \Ni{3}{k_0 {\bf r}_j}{pmn} 
+
\uTE{\bar pmnl}{j} \inoutTE{nl}{j} \Mi{3}{k_0 {\bf r}_j}{\bar pmn}
\right]&{\bf r}_j\in \Omega_{\overline{j}}\bigcup\Omega_3.
  \end{array} \right.
\label{eq:Single-sphere-scatt}
\end{equation}
where we have defined the following binary operator $(\bar{\cdot})$: $ \bar{1} = 2 $ and $\bar{2} = 1 $. 

The field incident on the $j$-th sphere is only the field produced by the remaining $\bar{j}$-th sphere. 
Therefore, by applying Eq. \ref{eq:DimerModeExpansion} of the main manuscript to the $j$-th sphere, we obtain:
\begin{equation}
\begin{aligned}
\uTM{pmnl}{j} &= \frac{\gamma -1}{\eigTMl_{nl}-\gamma}\,\, 
\Proj{TM}{\C{\bar{j}}_{pm}},\\
\uTE{pmnl}{j} &= \frac{\gamma -1}{\eigTEl_{nl}-\gamma}\,\, 
\Proj{TE}{\C{\bar{j}}_{pm}},
\end{aligned}
\label{eq:Projections}
\end{equation}
where $\Proj{TM}{\cdot},\Proj{TE}{\cdot}$ are defined in \ref{eq:ProjectionOnSingle}.\\
In order to apply \ref{eq:Projections}, and take full advantage of the orthogonality among VSWFs, we have to represent the field $\C{\bar{j}}_{pm}$ in the reference system $O_{\bar{j}}\rbar$. We now use the VSWF translation-theorem, which enables  us to represent the radiating VSWF centred at one origin, i.e. $\rbar$ as an expansion of regular VSWF centered about another origin $\rj$. It can be written as:
\begin{equation}
\label{eq:Translation_theorem1}
\begin{aligned}
\left( 
   \begin{array}{cc} \Mi{3}{k_0 {\bf r}_{\overline j}}{pmn} \\ \Ni{3}{k_0 {\bf r}_{\overline j}}{pmn}  \end{array} \right)
   =
   \displaystyle\sum_{\nu = \max\left(1,m \right)}^{\infty} 
\left[
\left(\begin{array}{cc}Q_{\mathtt{MM}_{mn\nu}}^{(j)}(d)\, \Mi{1}{k_0 {\bf r}_j}{pm\nu}\\Q_{\texttt{NN}_{mn\nu}}^{(j)}(d)\,\Ni{1}{k_0 {\bf r}_j}{pm\nu}\end{array}\right)
+ \left(\begin{array}{cc}Q_{\texttt{MN}_{pmn\nu}}^{(j)}(d)\, \Ni{1}{k_0 {\bf r}_j}{\overline pm\nu}\\ Q_{\texttt{NM}_{pmn\nu}}^{(j)}(d)\, \Mi{1}{k_0 {\bf r}_j}{\overline pm\nu}\end{array}\right)\right]\,
\end{aligned}
\end{equation}
where the translation-addition coefficients are given by \cite{xu95,xu96}:
\begin{equation}\label{eq:Translation_theorem2}
\begin{aligned}
Q_{\texttt{MM}_{m \nu n}}^{(j)}(d)&=Q_{\texttt{NN}_{m \nu n}}^{(j)}(d)&=  \frac{1}{2} \left[ A_{m\nu}^{mn}(d,\overline j\rightarrow j)+ \frac{\Gamma_{m\nu}}{\Gamma_{mn}}A_{-m\nu}^{-mn}(d,\overline j\rightarrow j)\right],\\
Q_{\texttt{MN}_{\substack{o \\ e} m \nu n}}^{(j)}(d)&=Q_{\texttt{NM}_{\substack{o \\ e} m \nu n}}^{(j)}(d)&=\pm \frac{i}{2}\left[ B_{m\nu}^{mn}(d,\overline j\rightarrow j )- \frac{\Gamma_{m\nu}}{\Gamma_{mn}}B_{-m\nu}^{-mn}(d,\overline j\rightarrow j)\right],\\
\end{aligned}
\end{equation}
\begin{equation}\label{eq:Translation_theorem3}
\begin{aligned}
&\Gamma_{mn}=(-1)^m \frac{(n-m)!}{(n+m)!},\\
&A_{m\nu}^{mn}(d,\overline j\rightarrow j) = \left\{
\begin{array}{ll}
(-1)^{n+\nu}\displaystyle\frac{E_{m\nu}}{E_{mn}} C_0\displaystyle\sum_{q=0}^{\min{\left(n,\nu\right)}} i^{p}\,C_p\, a_{-m n \nu m q} \,h_{p}^{(1)}(d) \quad &j=1,\\
\displaystyle\frac{E_{m\nu}}{E_{mn}} C_0\displaystyle\sum_{q=0}^{\min{\left(n,\nu\right)}} i^{p}\,C_p\, a_{-m n \nu m q} \,h_{p}^{(1)}(d) &j=2,
\end{array}\right.\\
&B_{m\nu}^{mn}(d,\overline j\rightarrow j) = \left\{
\begin{array}{ll}
(-1)^{n+\nu}\displaystyle\frac{E_{m\nu}}{E_{mn}} C_0\displaystyle\sum_{q=0}^{\min{\left(n,\nu\right)}} i^{p+1}\,C_p\, b_{m n \nu m q}\, h_{p+1}^{(1)}(d) \quad &j=1,\\
\displaystyle\frac{E_{m\nu}}{E_{mn}} C_0\displaystyle\sum_{q=0}^{\min{\left(n,\nu\right)}} i^{p+1}\,C_p\, b_{m n \nu m q}\, h_{p+1}^{(1)}(d) \quad &j=2,
\end{array}\right.\\
&p=n+\nu - 2q,\\
&E_{mn}=i^n \sqrt{\frac{(2 n+1) (n-m)!}{n (n+1) (m+n)!}},\\
&C_0=\frac{1}{2} (-1)^m \sqrt{\frac{(2 \nu +1) (2 n+1) (\nu -m)! (m+n)!}{\nu  (\nu +1) n (n+1) (m+\nu )! (n-m)!}},\\
&C_p = (\nu +1) \nu +(n+1) n-p (p+1),
\end{aligned}
\end{equation}
$a_{mn\nu m q}$ are the Gaunt coefficients and $b_{mn\nu m q}$ are combinations of the Gaunt coefficients, whose expression can be found in Refs. \cite{xu95,xu96}. \\
By substituting Eq. \ref{eq:Single-sphere-scatt} into Eq. \ref{eq:Projections} and by using the Eq. \ref{eq:Translation_theorem1}-\ref{eq:Translation_theorem3}, and truncating the summation indices $n$ and $l$  to the values $\Nmax$ and $\Lmax$, we obtain two coupled sets of homogeneous  equations for any given pair of $m \in \mathbb{N}_0$ and $p = e,o$ 
\begin{multline}
\frac{1}{\eigTMl_{nl}-1}\left\{
\uTM{pmnl}{j}+\Proj{TM}{\Cah{pmnl}} \sum_{\nu = \max{\left( 1,m \right)}}^{N_{max}}  \sum_{s=1}^{L_{max}} \left[
\uTM{pmnl}{\overline{j}} \inoutTM{\nu s}{\overline{j}}
Q_{\texttt{NN}_{m \nu n}}^{(j)}+ \uTE{\overline pm \nu s}{\overline{j}}
 \inoutTE{\nu s}{\overline{j}}
Q_{\texttt{MN}_{p m \nu n}}^{(j)}  \right]\right\} \\ = \frac{1}{\chi} \uTM{pmnl}{j}
\label{eq:systemEq1}
\end{multline}
\begin{multline}
 \frac{1}{\eigTEl_{nl}-1} \left\{
 \uTE{\overline pm n l}{j}  + \Proj{TE}{\Cbh{pmnl}} \sum_{\nu = \max{ \left(1,m\right)}}^{N_{max}} \sum_{s=1}^{L_{max}}\left[
\uTE{\overline pm\nu s}{\overline{j}} \inoutTE{\nu s}{\overline{j}} Q_{\texttt{MM}_{m \nu n}}^{(j)}
+\uTM{pm\nu s}{\overline{j}} 
\inoutTM{\nu s}{\overline{j}}
Q_{\texttt{NM}_{\overline pm \nu n}}^{(j)}
 \right] \right\} \\ = \frac{1}{\chi} \uTE{\overline pmnl}{j}
\qquad \mbox{with}
\left(
\begin{array}{c}
n = \max\left( 1, m \right), \dots, N_{max} \\
\quad l = 1, \ldots, L_{max} \\
j=1,2
\end{array}\right)
\label{eq:systemEq2}
\end{multline}
where $\chi = \gamma -1$ is the  eigen-susceptibility of the two sphere. For any given $m$ and $p$ indices, we have a system of $Q_{max} = \left[ 4 \Lmax \left( \Nmax  - \max{\left(1,m\right)} + 1 \right) \right]$ equations.
It can be written in the matrix form:
\begin{equation}
 {\bf A}^{(m,p)} {\bf y}^{(m,p)} = \xi^{(m,p)} \, {\bf y}^{(m,p)}
\end{equation}
where ${\bf y}_{pm}= \left[ \uTM{pmnl}{j}, \,\uTE{pmnl}{j} \right]^T$ is the vector containing the expansion coefficients. We numerically evaluate the finite number of eigenvectors of the matrix $ {\bf A}^{(m,p)}$.

For any pair of indices $p,m$, we have the eigenvalues
$\chi_q^{pm}$ for  $q=1,\ldots,  Q_{max}$.   Starting from the eigenvalues $\chi_q^{pm}$, it is possible to obtain the dimer eigen-permittivities  $\EigDimer{pmq}$ through the relation: $\EigDimer{pmq} = 1/\chi_q^{pm}+1$. The $q$-th eigenvectors of the discrete problem is denoted as $\vecD$, its coefficients are $[\uTM{pmnlq}{j},\,\,\uTE{pmnlq}{j}]$

The electric-field modes can be obtained from the coefficient eigenvector $\vecD$ by using Eqs. \ref{eq:Cdimer}-\ref{eq:Single-sphere-scatt}
\begin{equation}
    \CC{pmq} =  \sum_{j=1}^2 {\bf C}_{pmq}^{\left(j\right)}
\label{eq:Eig-Single-sphere-scatt1Final},
\end{equation}
\begin{equation}
{\bf C}_{pmq}^{\left(j\right)} \left( {\bf r}_j \right) =
\displaystyle\sum_{ n l}
\left[
\uTM{pmq \, nl}{j} \Cah{pmnl}({\bf r}_j)
+
\uTE{\bar{p}mq \, nl}{j} \Cbh{\bar{p}mnl}({\bf r}_j)
\right].
\label{eq:Single-sphere-scatt1Final}
\end{equation}
In Eq. \ref{eq:Eig-Single-sphere-scatt1Final} the mode $\CC{pmq}$ within each sphere is represented in terms of the superposition of two contributions ${\bf C}_{pmq}^{\left(1\right)}$ and ${\bf C}_{pmq}^{\left(2\right)}$ centred in different reference systems. In other words, the dimer mode within each sphere is represented in terms of the isolated sphere modes of the sphere $1$ {\it and} of the sphere $2$.
Nevertheless, it is possible to overcome this problem, by representing the dimer electric field mode $\CC{pmq}$ within the $j$-th sphere exclusively in terms of the isolated sphere modes $\Cah{pmnl}$ and $\Cbh{pmnl}$ by using the translation-addition theorem for vector spherical wave functions (VSWF). Thus, we obtain the following representation of $\CC{pmq}$:
\begin{eqnarray}
\label{eq:Dimer_Mode_Expansion}
\CC{pmq}({\bf r}_j)=\ \left\{
\begin{array}{lll}
&\displaystyle\sum_{n   l} \CoeffdTMmet{pmq \, nl} \Ni{1}{\SqrtTM \, k_0 {\bf r}_j}{pmn} 
+ \CoeffdTEmet{\overline pmq \, nl} \Mi{1}{\SqrtTE \, k_0  {\bf r}_j}{\overline pmn}, \qquad&{\bf r}_j\in \Omega_j\\
&\displaystyle\sum_{j=1}^2\displaystyle\sum_{n    l} \, \uTM {pmq\,nl}{j} \inoutTM{nl}{j} \, \Ni{3}{k_0 {\bf r}_j}{pmn}
+ \uTE {\overline pmq\, nl}{j} \inoutTE{nl}{j} \Mi{3}{ k_0 {\bf r}_j}{\overline pmn}, &{\bf r}_j\in \Omega_3
\end{array}
\right.
\end{eqnarray}
where the coefficients $\{ \CoeffdTMmet{pmq\,nl} ,\, \CoeffdTEmet{pmq\,nl}  \}$ are:
\begin{equation}
\begin{aligned}
\CoeffdTMmet{pmq\,nl} &= \uTM{pmq\,nl}{j}+
 \Proj{TM}{\Cah{pm \, nl}}\sum_{\nu=max{\left(1,m \right)}}^{N_{max}} \sum_{s=1}^{L_{max}}\left[ 
 \uTM{pmq \, \nu s}{\overline j} \inoutTM{\nu s}{\overline{j}}
Q_{\texttt{NN}_{m \nu n}}^{(j)}+
\uTE{\overline pmq\,\nu s}{\overline j} \inoutTE{\nu s}{\overline{j}} 
Q_{\texttt{MN}_{p m \nu n}}^{(j)} \right]\\
\CoeffdTEmet{pmq\,nl} &= \uTE{pmq\,nl}{j} +
\Proj{TE}{\Cbh{pm\,nl}}\sum_{\nu=max{\left(1,m \right)}}^{N_{max}} \sum_{s=1}^{L_{max}}\left[
 \uTE{ pmq\,\nu s}{\overline j} \inoutTE{\nu s}{\overline{j}}
Q_{\texttt{MM}_{m \nu n}}^{(j)}
+ \uTM{\overline p mq \, \nu s}{\overline j}
 \inoutTM{\nu s}{\overline{j}}
Q_{\texttt{NM}_{ pm \, \nu n}}^{(j)}
 \right]
\end{aligned}
\end{equation}
